\documentclass[aps,prapplied,twocolumn,reprint,nofootinbib,longbibliography]{revtex4-2}
\usepackage[left=1.5cm,right=1.5cm,top=2.5cm,bottom=2.5cm]{geometry}
\usepackage{graphicx}
\usepackage{subcaption}
\usepackage{booktabs}
\usepackage{amsmath,amssymb,bm}
\usepackage{dcolumn}
\usepackage[colorlinks=true,linkcolor=blue,citecolor=blue,urlcolor=blue]{hyperref}
\usepackage{xcolor}

\newcommand{\Bz}{B_z}
\newcommand{\By}{B_y}
\newcommand{\Bc}{B_{c}}
\newcommand{\muT}{\mathrm{\mu T}}
\newcommand{\Btot}{\ensuremath{B_z^{\mathrm{total}}}}
\newcommand{\musim}{I_{\mathrm{sim}}}
\definecolor{todocol}{rgb}{0.0,0.18,0.70}

\begin{document}

\title{Offline Evaluation of Information-Theoretic Adaptive Bayesian Vector Magnetometry}

\author{Dinesh Ramachandran}
\altaffiliation{These authors contributed equally to this work.}
% \affiliation{Department of Physics, IIT Kanpur, Kanpur 208016, India}
\author{Shubham Jaiswal}
\altaffiliation{These authors contributed equally to this work.}
% \affiliation{Department of Physics, IIT Kanpur, Kanpur 208016, India}
\author{Goutam Manna}
% \affiliation{Department of Physics, IIT Kanpur, Kanpur 208016, India}
\author{Saikat Ghosh}
\email{gsaikat@iitk.ac.in}

\affiliation{Department of Physics, IIT Kanpur, Kanpur 208016, India}

\date{\today}

\begin{abstract}
An information-theoretic adaptive Bayesian framework for jointly estimating the longitudinal and transverse components of a magnetic field from the time-resolved optical response of a cold $^{87}$Rb ensemble is presented. 
A linearly polarized continuous-wave probe drives the $D_2$ $F{=}1 \rightarrow F'{=}0$ transition, producing coherent population trapping (CPT) lineshapes whose width, contrast, and transient dynamics depend sensitively on the ambient magnetic field. 
The full temporal signal, including early-time transient signal and the steady-state absorption, is compared against a precomputed Maxwell-Bloch forward model. 
The joint posterior distributions of the longitudinal and transverse components of the magnetic field are updated sequentially, while marginalizing over instrumental nuisance parameters arising from the finite resolution of the magnetic bias coil supply. 
At each step, the applied longitudinal bias field is chosen to maximize the expected Kullback-Leibler (KL) divergence between the current prior and the anticipated posterior, the expected information gain. 
As a proof of principle, the protocol is validated in post-processing on pre-recorded experimental scans, and both field components converge to well localized field estimates with weak correlation between them, consistent with the distinct dependence of the CPT resonance position and lineshape on the field components. 
These results establish an information-theoretic framework for joint vector field estimation from transient cold-atom spectroscopy. However, there exists
a systematic forward-model discrepancy, showing that posterior concentration alone does not establish accuracy and motivating nuisance-parameter inference in future online implementations.
\end{abstract}

\maketitle

\section{Introduction}
\label{sec:intro}

Precision measurements of magnetic fields matter across physics, from fundamental symmetry tests to biomedical imaging and geophysical surveying \cite{BudkerRomalis2007,Degen2017}. 
Atomic magnetometers, which transduce magnetic information into optical signals through spin-dependent atom-light interactions, have reached sensitivities rivaling and in some regimes surpassing those of superconducting quantum interference devices \cite{BudkerRomalis2007,Kominis2003}.

The physics of these magneto-optical effects is well described by density-matrix (Maxwell-Bloch) models of multilevel atoms interacting with near-resonant light in the presence of external magnetic fields \cite{Fleischhauer2005,Arimondo1996,Auzinsh2010}. 
For the $D_2$ line of $^{87}$Rb, the $F{=}1 \rightarrow F'{=}0$ transition realizes a clean $\Lambda$-type system when driven by linearly polarized light: the two ground-state magnetic sublevels $|F{=}1, m_F{=}{\pm}1\rangle$ couple via $\Omega_p^{+}$ and $\Omega_p^{-}$ circular polarization components to a single excited state, forming a dark superposition whose properties depend sensitively on the magnetic field \cite{Arimondo1996,Wynands1999,Rochester2001}. 
The atomic parameters of $^{87}$Rb are precisely tabulated by Steck \cite{Steck}. 
Dark resonances in this system appear as narrow CPT features in the transmitted light, with a width, contrast, and dispersive profile that vary with both the magnitude and the direction of the ambient field \cite{Budker2002,Pustelny2008,Wynands1999}. 
Related coherent resonances driven by linearly polarized light have also been used for vector-sensitive magnetometry, demonstrating that the resonance response can encode directional information about the magnetic field \cite{Yudin2010}. 
The optical Bloch equations for this system admit both steady-state and time-domain solutions, and the transient dynamics, including optical nutation, dark-state pumping, and coherence buildup, carry information that the steady-state lineshape alone does not \cite{Fleischhauer2005,Auzinsh2010}.

Despite this directional sensitivity, most atomic magnetometers remain single-axis devices, and extracting multiple components of the magnetic field generally requires additional experimental complexity. 
Demonstrated approaches include auxiliary modulation fields \cite{Seltzer2004,Patton2014,LeGal2021}, multiple probe beams or electromagnetically induced transparency resonances \cite{Afach2015,Cox2011}, the Voigt effect \cite{Pyragius2019}, microwave polarization reconstruction \cite{Thiele2018}, and fast-rotating fields applied to scalar sensors \cite{Wang2025}. 
Vector magnetometry nevertheless remains more demanding than scalar measurement: heading errors, dead zones, and cross-coupling between components require careful calibration and control of systematics \cite{Patton2014,Wang2025}.

Bayesian inference provides a convenient framework for extracting parameters from nonlinear forward models while retaining a joint probability distribution over the unknown quantities and allowing uncertain experimental parameters to be treated as nuisance variables \cite{vonderlinden2014,Granade2012}. 
The inference can further be combined with adaptive experimental design, in which the next measurement setting is chosen to optimize the information it is expected to provide. 
A standard information-theoretic criterion is the expected Kullback-Leibler divergence, or expected information gain, introduced in Bayesian experimental design by Lindley and developed more generally in subsequent work \cite{Lindley1956,Chaloner1995}. 
Adaptive Bayesian strategies have been applied across quantum characterization and metrology, including quantum tomography and parameter estimation \cite{Huszar2012,Glatthard2022,Overton2026}. 
In magnetometry specifically, Bayesian inference and sequential experimental design have been used to improve parameter estimation and measurement efficiency in platforms including solid-state spin sensors and atomic magnetometers \cite{Dushenko2020,Puebla2021,Ma2025}. 
These works establish the utility of Bayesian inference and adaptive measurement selection in magnetometry; the present work does not seek to demonstrate a new adaptive performance figure of merit. 
% Instead, we use pre-recorded experimental data to test whether the transient CPT response of a cold-atom system can support joint estimation of multiple magnetic-field components together with posterior-dependent selection of the longitudinal bias field.
Instead, pre-recorded experimental data is used with posterior-dependent selection of the longitudinal bias field to test whether the information gained about the components of the ambient magnetic field can be optimized. 

Measurement-based feedback also has a distinct history in quantum control, where information obtained from a quantum system is used to modify its subsequent quantum evolution or stabilize a desired quantum state \cite{Wiseman_Milburn_2009}. 
The feedback considered here is not of this type. 
The inferred magnetic-field posterior is used only to select the experimental control setting for a subsequent measurement realization; no feedback is applied to steer or stabilize the quantum state of the ensemble currently being measured. 
The resulting loop is therefore one of classical adaptive experimental design applied to a quantum sensor. 
More general feedback protocols could, in principle, vary the control parameter during a single atomic evolution. 
Such a scheme would require a forward model that explicitly accounts for the resulting time-dependent atomic dynamics and the history of the applied control, in addition to the associated experimental constraints. 
The present work does not consider this case. 
Instead, each selected control is held fixed during an individual measurement realization.

This work combines cold-atom magneto-optical spectroscopy, density-matrix forward modeling, joint Bayesian field estimation, and information-theoretic adaptive design in an offline proof-of-concept study. 
A cold $^{87}$Rb ensemble in a MOT is probed with linearly polarized continuous-wave light on the $D_2$ $F{=}1 \rightarrow F'{=}0$ transition and the CPT response as a function of an applied longitudinal bias field is recorded. 
A density-matrix forward model computes the full time-resolved optical response, including both the transient dynamics during optical pumping and the steady-state signal.

The ambient longitudinal and transverse magnetic-field components, $(B_z,B_y)$, are inferred simultaneously through a joint Bayesian posterior, while the applied longitudinal bias field is selected using an expected information gain criterion. 
The method is evaluated entirely in post-processing using pre-recorded experimental scans. 
The purpose of this offline analysis is to test, directly on archived experimental data, whether the measured transient response, together with the available forward model and information-guided measurement selection, supports joint field estimation before introducing the additional requirements of real-time control. 
% We therefore do not claim here an experimentally demonstrated improvement in sensitivity, dynamic range, measurement time, or data efficiency relative to non-adaptive protocols. 
% Such figures of merit require a controlled online implementation with a well-defined experimental acquisition budget.

The consistency of the inferred fields across independently calibrated transverse field settings is further examined and using likelihood scale studies, model-data residuals, and full-scan consistency tests the limitations imposed by the present forward model are characterized. 
The resulting analysis identifies both the promise of the joint estimation scheme and the model-discrepancy and nuisance-parameter issues that must be addressed in a future online implementation.

The remainder of this paper is organized as follows. 
Section~\ref{sec:setup} describes the experimental setup and the density-matrix forward model. 
Section~\ref{sec:protocol} details the adaptive estimation protocol, including the joint Bayesian update and the expected KL divergence utility function. 
Section~\ref{sec:results} presents the proof-of-concept results. 
Section~\ref{sec:discussion} discusses the limitations of the present forward model and the requirements for future online implementation, and Sec.~\ref{sec:conclusion} concludes with an outlook.

\section{Experimental setup and forward model}
\label{sec:setup}

\subsection{Cold-atom Platform}
\label{sec:setup_cold_atom_platform}

The experiments are performed on a cold ensemble of $^{87}$Rb atoms confined in a magneto-optical trap, with a typical post-cooling temperature of ${\approx}100~\mu$K \cite{Laskar2020}. The cloud is nearly ellipsoidal, with a length of $4.1$~mm along the probe direction; the probe beam diameter of $1.5$~mm sets the sensing volume of the ensemble. In each experimental cycle, optical pumping prepares roughly $4\times10^{6}$ atoms in the $F=1$ hyperfine level before probing. Three pairs of Helmholtz coils apply and control the magnetic field components along $x$, $y$, and $z$; the maximum bias field along $z$ is $\pm340~ \mathrm{\mu T}$.

\begin{figure}[t]
\includegraphics[width=0.9\columnwidth]{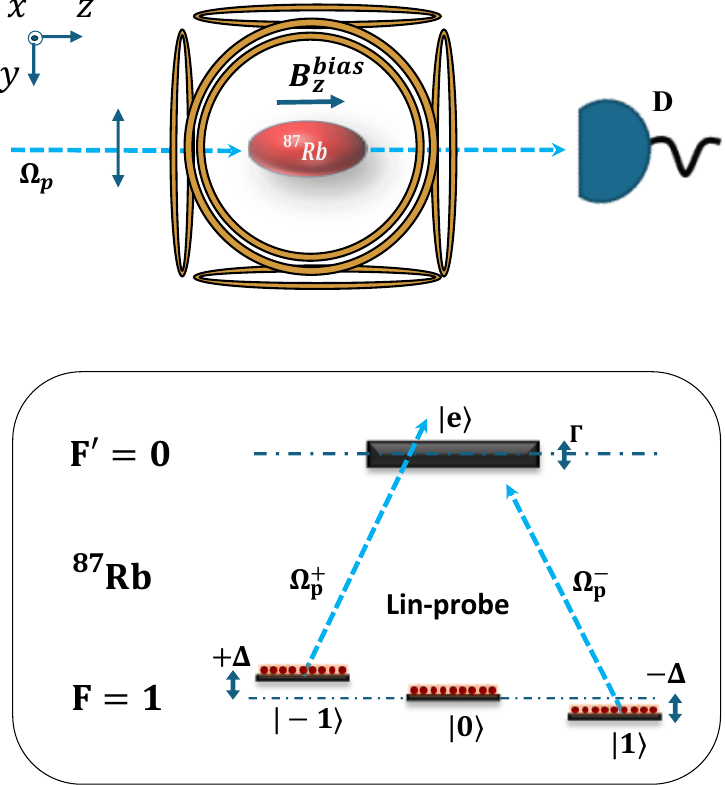}
% \vspace{5mm}
\caption{Experimental schematic. A linearly polarized probe interacts with a cold $^{87}$Rb ensemble along the $z$ axis. The applied longitudinal bias field $B_z^{\mathrm{bias}}$ is the adaptive control variable. Inset: the $D_2$ $F{=}1 \rightarrow F'{=}0$ level structure showing $\Omega_p^{+}/\Omega_p^{-}$ coupling to a single excited state, forming a $\Lambda$-type CPT system. The ground-state Zeeman sublevels are shifted by the Larmor detuning ($\pm\Delta$).}
\label{fig:setup}
\end{figure}

A linearly polarized probe beam, polarized along $y$ and tuned near the $D_2$ $F{=}1 \rightarrow F'{=}0$ transition at $\lambda = 780$~nm, passes through the atomic cloud along the $z$ axis (Fig.~\ref{fig:setup}). 
In the atomic frame, the linear polarization decomposes into equal $\Omega_p^{+}$ and $\Omega_p^{-}$ circular components, coupling the $|F{=}1, m_F{=}{-}1\rangle$ and $|F{=}1, m_F{=}{+}1\rangle$ ground-state sublevels to the single excited state $|F'{=}0, m_{F'}{=}0\rangle$. 
This realizes a textbook $\Lambda$ system in which a dark superposition of the two ground states decouples from the light field \cite{Fleischhauer2005,Arimondo1996}. 
After interacting with the atoms, the transmitted light passes through a Glan-Thompson analyzer with its pass axis also along $y$, and serves as the measurement signal. The observed fluctuations in the signal comprise a dominant electronic noise contribution from the detection and acquisition chain together with an additional optical intensity dependent contribution. The resulting measurement noise was characterized experimentally under the relevant operating conditions. Its relation to the effective likelihood scale used for
inference, and the dependence of the inferred fields on that scale, are examined in Sec.~\ref{sec:sigma_dependence}.
% \TODO{(Done) Add details about the noise in the experiment, and what order we have estimated its value to be.}
The externally applied longitudinal field $B_z^{\mathrm{bias}}$ is the control parameter ($B_c$), while the ambient transverse field $B_y$ and any residual longitudinal offset $B_z$ are the target parameters to be estimated.

In the present analysis, the Bayesian estimation is restricted to the $B_y$ and $B_z$ components. The $B_x$ component is independently compensated and is therefore not included as a parameter being estimated. A small residual $B_x$ may nevertheless remain after the compensation; its effect could be incorporated in a generalized Bayesian framework by treating $B_x$ as a nuisance parameter marginalizing over its uncertainty. 
% We do not attempt to estimate the transverse component $B_x$ in this work, for two reasons. 
% First, with the probe propagating along $z$ and polarized along $y$, the residual ambient field along $x$ in our apparatus is cancelled, therefore enters the lineshape only weakly relative to $B_y$; its principal effect is a modest reduction of the CPT contrast, which is absorbed into the calibration of the forward model. 
Estimating $B_x$ would also require extending the precomputed simulation grid to a third field dimension, substantially increasing the computational cost of both the grid generation and the likelihood evaluation, which is not warranted for the present offline demonstration. 
The consequences of this choice, and the route to a full three-axis estimate, are discussed in Sec.~\ref{sec:discussion}. Further, because the signal is even in $B_y$, the inference is restricted to $B_y \geq 0$. Use of a control bias field along $y$ will help break this degeneracy, but that remains part of future work.

The archived datasets acquired are as follows: for each independently calibrated transverse field setting measured by Free Induction Decay (FID), \(B_{y,\mathrm{FID}}\), the time-resolved probe transmission over a discrete scan of longitudinal bias fields \(B_c\) was recorded. Each bias-field scan corresponds to an independently prepared atomic ensemble for which \(B_c\) is applied before probing and held fixed throughout the recorded transient. The resulting archived dataset therefore forms a measured response surface \(I_{\mathrm{exp}}(t,B_c;B_{y,\mathrm{FID}})\). Four such datasets, acquired at different calibrated transverse fields, are used in this work; one is used as a representative case for illustrating the adaptive inference (Sec.~\ref{sec:estimate_convergence}), while all four are used to test consistency across transverse field settings (Sec.~\ref{sec:transverse_field_consistency}).

\subsection{Density-matrix Forward Model}
\label{sec:forward}

The theoretical prediction for the transmitted probe signal is generated by numerically solving the Maxwell-Bloch equations for the four-level system in the presence of an arbitrary magnetic field \cite{Fleischhauer2005,Auzinsh2010}. 
The density matrix $\rho(t)$ evolves under the Liouville-von Neumann equation
\begin{equation}
\frac{d\rho}{dt} = -\frac{i}{\hbar}\,[H, \rho] + \mathcal{L}[\rho]\,,
\label{eq:liouville}
\end{equation}
where the Hamiltonian $H$ includes the atom-light interaction in the rotating-wave approximation and the Zeeman shifts due to the total effective magnetic field $\mathbf{B}_{\mathrm{eff}} = (B_x, B_y, B_z + \Bc)$, with $\Bc \equiv B_z^{\mathrm{bias}}$ denoting the applied control field. 
The Lindblad superoperator $\mathcal{L}$ accounts for spontaneous emission from $|F'{=}0\rangle$ at the rate $\Gamma$. The ground-state Zeeman Hamiltonian is
\begin{equation}
H_{\mathrm{Z}} = g_F \mu_B \sum_q B_q F_q\,,
\label{eq:zeeman}
\end{equation}
where $g_F$ is the Land\'e $g$ factor of the $F{=}1$ ground state, $\mu_B$ is the Bohr magneton, and $F_q$ denotes the cartesian components of the angular-momentum operator \cite{Steck}. 
The transverse components $B_x$ and $B_y$ mix the $m_F = 0, \pm 1$ sublevels, modifying the dark-state composition, the CPT resonance width and contrast, and the transient approach to steady-state.

The transmitted probe intensity $I_{sim}(t, B_z, B_y, \Bc)$ is obtained from the coherence and population elements of $\rho(t)$ through the susceptibility of the medium. 
To enable fast, repeated evaluation during the Bayesian update, $\musim$ is precomputed on a discrete three-dimensional grid spanning $(t, \Btot, B_y)$ and accessed via linear interpolation, where $\Btot = \Bz + \Bc$ is the total effective longitudinal field during a given measurement. 
The experimental and simulated probe intensities are expressed in terms of saturation intensity ($I_{\mathrm{sat}}$) of the rubidium transition \cite{Steck}.
% \TODO{Add details about $I_{sat}$. This statement is phrased misleadingly. We did not calibrate Isat, we know Isat, and the response is calibrated in terms of Isat. and refer to daniel steck. How?}

In the simulation, the longitudinal field $\Btot$ spans $-344.6$ to $344.6~\muT$ with a grid resolution of ${\approx}7.66~\muT$. 
The transverse field $B_x$ is held constant at $0~\muT$, and $B_y$ is varied from $0.0$ to $71.4~\muT$ with a grid resolution of ${\approx}1.4~\muT$. 
The probe Rabi frequency is cross-referenced with the measured probe intensity and fixed at $\Omega_p = 0.0792\,\Gamma$. 
The ground-state and thermal decoherence rates are fixed at $15$~kHz and $1$~kHz, respectively. These values are determined independently from the data used for Bayesian estimation and are chosen to reproduce the experimentally observed resonance contrast and full width at half maximum (FWHM) of the magnetic resonance signal.

\section{Adaptive Bayesian Protocol}
\label{sec:protocol}

\subsection{Measurement Model}
\label{sec:measurement_model}

The parameters of interest are the ambient longitudinal and transverse
magnetic-field components, assumed constant throughout the estimation process,
\begin{equation}
    \mathbf{B}
    \equiv
    \begin{pmatrix}
        \Bz \\
        \By
    \end{pmatrix},
    \label{eq:magnetic_field_vector}
\end{equation}
while the externally applied longitudinal bias field $\Bc$ serves as the experimental control variable.

In the intended online adaptive experiment, the $k^{\mathrm{th}}$ Bayesian update corresponds to an independent experimental realization, where a fresh atomic ensemble is prepared, a fixed bias field of our choice, $B_c^{(k)}$ is applied and the complete transient probe is recorded. The total field is hence constant throughout that realization and is only changed between successive realizations.

The experimentally observed probe
transmission is represented by
\begin{equation}
    \mathbf{y}_{\mathrm{exp}}^{(k)}
    =
    \begin{bmatrix}
        y^{(k)}(t_1) &
        y^{(k)}(t_2) &
        \cdots &
        y^{(k)}(t_N)
    \end{bmatrix}^{\mathsf T},
\end{equation}
\(t_n\) ($n=1,2,...,N$) denotes the time measured from the beginning of the probe. The
corresponding prediction from the density-matrix forward model is
\begin{multline}
    \boldsymbol{I}^{(k)}_{\mathrm{sim}}
    \left(
        \mathbf{B},
        B_c^{(k)}
    \right)
    = \\
    \begin{bmatrix}
        \musim^{(k)}
        \left(
            t_1,\mathbf{B},B_c^{(k)}
        \right)
        &
        \cdots
        &
        \musim^{(k)}
        \left(
            t_N,\mathbf{B},B_c^{(k)}
        \right)
    \end{bmatrix}^{\mathsf T}.
\end{multline}
An independent Gaussian residual model is adopted, with
\begin{multline}
    y^{(k)}(t_n)
    =
    \musim^{(k)}
    \left(
        t_n,\mathbf{B},B_c^{(k)}
    \right)
    +
    \epsilon_n,
    \qquad
    \epsilon_n
    \sim
    \mathcal{N}(0,\sigma^2),
\end{multline}
which gives the conditional likelihood
\begin{multline}
    p\left(
        \mathbf{y}_{\mathrm{exp}}^{(k)}
        \mid
        \mathbf{B},B_c^{(k)}
    \right)
    = \\
    \frac{1}{(2\pi\sigma^2)^{N/2}}
    \exp\left[
        -\frac{
            \left\|
                \mathbf{y}_{\mathrm{exp}}^{(k)}
                -
                \boldsymbol{I}^{(k)}_{\mathrm{sim}}
                \left(
                    \mathbf{B},B_c^{(k)}
                \right)
            \right\|_2^2
        }{2\sigma^2}
    \right].
\end{multline}
%
% Equivalently, the log-likelihood is
% %
% \begin{multline}
%     \log
%     p\left(
%         % \mathbf{y}_{\mathrm{exp}}^{(k)}
%         \mid
%         \mathbf{B},B_c^{(k)}
%     \right)
%     = \\
%     -
%     \frac{1}{2\sigma^2}
%     \left\|
%         \mathbf{y}_{\mathrm{exp}}^{(k)}
%         -
%         \boldsymbol{\mu}^{(k)}_{\mathrm{sim}}
%         \left(
%             \mathbf{B},B_c^{(k)}
%         \right)
%     \right\|_2^2
%     + C,
%     \label{eq:joint_log_likelihood}
% \end{multline}
% %
% where \(C\) is independent of \(\mathbf{B}\).

The likelihood scale parameter $\sigma$ is expressed in the same $I_{\mathrm{sat}}$ units as the experimental and simulated probe signals. 
The experimental intensity noise was independently characterized and is of the order $\sim 10 ^{-4} \,I_{\mathrm{sat}}$.
In the likelihood it is treated as
an effective confidence scale rather than being fixed directly to the measured
detector noise level. Its influence on the inferred fields is examined in
Sec.~\ref{sec:sigma_dependence}.

% In the intended online experimental realization, as stated earlier, one measurement would
% consist of preparing the atomic ensemble, optically pumping it, applying a single fixed value of $\Bc$, and recording the complete transient response. 
% The posterior obtained from that realization would then determine the bias field $\Bc^{(k+1)}$ used for the next independently prepared
% ensemble.

% The present work instead evaluates this measurement-selection policy in post-processing using pre-recorded experimental scans. 
% Successive Bayesian updates therefore ingest temporal blocks of duration \(\Delta t_{\rm step} = 0.2~\mu\mathrm{s}\) from the archived time-series data at the bias field functioning as an adaptive search. 
% This temporal partitioning provides a computational means of examining the evolution of the joint posterior and the associated information-theoretic search policy while possessing the same forward-model as the online case and 
% it should not be interpreted as changing the applied bias field during the physical evolution of a single atomic realization.

\subsection{Bayes' Theorem Update}
\label{sec:joint_bayesian_update}

At the $k^{\mathrm{th}}$ update, the unknown magnetic-field vector components are inferred
simultaneously through the joint posterior
\begin{equation}
    p\left(
        \mathbf{B}
        \mid
        \mathbf{y}_{1:k}
    \right),
\end{equation}
where
\(\mathbf{y}_{1:k}
=
\{\mathbf{y}_{\mathrm{exp}}^{(1)},\ldots,
\mathbf{y}_{\mathrm{exp}}^{(k)}\}\)
denotes all data incorporated up to the current update and $\mathbf{B}$ is as defined in Eq.~(\ref{eq:magnetic_field_vector}).

For a perfectly known applied bias field \(B_c^{(k)}\), Bayes' rule gives
\begin{equation}
    p\left(
        \mathbf{B}
        \mid
        \mathbf{y}_{1:k},
        B_c^{(k)}
    \right)
    =
    \frac{
        p\left(
            \mathbf{y}_{\mathrm{exp}}^{(k)}
            \mid
            \mathbf{B},B_c^{(k)}
        \right)
        p\left(
            \mathbf{B}
            \mid
            \mathbf{y}_{1:k-1}
        \right)
    }{
        p\left(
            \mathbf{y}_{\mathrm{exp}}^{(k)}
            \mid
            \mathbf{y}_{1:k-1},B_c^{(k)}
        \right)
    }.
    \label{eq:joint_bayes_rule_exact_bias}
\end{equation}
The denominator is the Bayesian evidence and ensures normalization over the
joint \((B_z,B_y)\) parameter space.

In the experiment, however, the bias field applied by the coil is not known
exactly. The control algorithm requests a set-point
$B_{c,*}^{(k)}$, while the actual field can differ from this value because
of the finite resolution of the coil-current supply.
Therefore, the realized bias field $B_c$ is treated as an instrumental nuisance
parameter rather than fixing it to its chosen value.

The uncertainty in the applied bias is represented by the distribution
\begin{equation}
    \pi_c
    \left(
        B_c
        \mid
        B_{c,*}^{(k)}
    \right)
    =
    \mathcal{U}
    \left(
        B_{c,*}^{(k)}- \delta B_c,
        B_{c,*}^{(k)}+ \delta B_c
    \right),
    \label{eq:bias_uncertainty_prior}
\end{equation}
where \(\delta B_c\) denotes the bias-field uncertainty associated with the coil supply, which in our post-processing context, would be the difference between adjacent set-points.

The likelihood in the Bayesian update is consequently obtained by marginalizing over all bias fields that could have been realized:
\begin{multline}
\begin{gathered}
    \widetilde{\mathcal{L}}_k(\mathbf{B})
    = p\left(
            \mathbf{y}_{\mathrm{exp}}^{(k)}
            \mid
            \mathbf{B},B_{c, *}^{(k)}
        \right) \\
    = \int
    p\left(
        \mathbf{y}_{\mathrm{exp}}^{(k)}
        \mid
        \mathbf{B},B_c
    \right)
    \pi_c
    \left(
        B_c
        \mid
        B_{c,*}^{(k)}
    \right)
    \,dB_c .
    \label{eq:bias_marginalized_likelihood}
\end{gathered}
\end{multline}
For the uniform distribution in Eq.~\eqref{eq:bias_uncertainty_prior}, this
becomes
\begin{equation}
    \widetilde{\mathcal{L}}_k(\mathbf{B})
    =
    \frac{1}{2 \delta B_c}
    \int_{B_{c,*}^{(k)}-\delta B_c}
         ^{B_{c,*}^{(k)}+\delta B_c}
    p\left(
        \mathbf{y}_{\mathrm{exp}}^{(k)}
        \mid
        \mathbf{B},B_c
    \right)
    \,dB_c .
    \label{eq:uniform_bias_marginalization}
\end{equation}
The integral is evaluated numerically using a discrete set of bias-field
values spanning the support of the instrumental distribution.

The joint posterior update is then
\begin{equation}
    p\left(
        \mathbf{B}
        \mid
        \mathbf{y}_{1:k}
    \right)
    =
    \frac{
        \widetilde{\mathcal{L}}_k(\mathbf{B})
        p\left(
            \mathbf{B}
            \mid
            \mathbf{y}_{1:k-1}
        \right)
    }{
        \displaystyle
        \int
        \widetilde{\mathcal{L}}_k(\mathbf{B}')
        p\left(
            \mathbf{B}'
            \mid
            \mathbf{y}_{1:k-1}
        \right)
        \,d\mathbf{B}'
    }.
    \label{eq:joint_bayes_update}
\end{equation}
Marginalizing over \(B_c\) is important because it propagates uncertainty in
the experimentally applied control field into the inferred magnetic-field
components. Fixing \(B_c=B_{c,*}^{(k)}\) would instead attribute discrepancies
arising from imperfect bias-field realization to \(B_z\), \(B_y\), or the
forward model, potentially producing overconfident or systematically shifted
posteriors.

% For numerical stability, the update is performed in logarithmic form:
% %
% \begin{equation}
%     \log \widetilde{p}_k(\mathbf{B})
%     =
%     \log
%     p\left(
%         \mathbf{B}
%         \mid
%         \mathbf{y}_{1:k-1}
%     \right)
%     +
%     \log
%     \widetilde{\mathcal{L}}_k(\mathbf{B}),
%     \label{eq:log_posterior_update}
% \end{equation}
% %
% after which the maximum log weight is subtracted before exponentiation and
% normalization.
The maximum-a-posteriori (MAP) estimate is
\begin{equation}
    \widehat{\mathbf{B}}_{\mathrm{MAP}}^{(k)}
    =
    \underset{\mathbf{B}}{\operatorname{arg\,max}}\;
    p\left(
        \mathbf{B}
        \mid
        \mathbf{y}_{1:k}
    \right).
    \label{eq:joint_map}
\end{equation}
The posterior mean and covariance matrix are evaluated as
\begin{equation}
    \boldsymbol{\mu}_{\mathbf{B}}^{(k)}
    =
    \int
    \mathbf{B}\,
    p\left(
        \mathbf{B}
        \mid
        \mathbf{y}_{1:k}
    \right)
    d\mathbf{B},
\end{equation}
and
\begin{equation}
    \boldsymbol{\Sigma}_{\mathbf{B}}^{(k)}
    =
    \int
    \left(
        \mathbf{B}
        -
        \boldsymbol{\mu}_{\mathbf{B}}^{(k)}
    \right)
    \left(
        \mathbf{B}
        -
        \boldsymbol{\mu}_{\mathbf{B}}^{(k)}
    \right)^{\mathsf T}
    p\left(
        \mathbf{B}
        \mid
        \mathbf{y}_{1:k}
    \right)
    d\mathbf{B}.
    \label{eq:posterior_covariance}
\end{equation}
Thus,
\begin{equation}
    \boldsymbol{\Sigma}_{\mathbf{B}}
    =
    \begin{pmatrix}
        \operatorname{Var}(B_z)
        &
        \operatorname{Cov}(B_z,B_y)
        \\
        \operatorname{Cov}(B_y,B_z)
        &
        \operatorname{Var}(B_y)
    \end{pmatrix},
\end{equation}
which quantifies both the marginal uncertainty in each field component and
the correlation between them.

\subsection{Posterior Grid and Adaptive Refinement}
\label{sec:posterior_grid_refinement}

At the $k^{\mathrm{th}}$ Bayesian update, the joint posterior is represented on the
two-dimensional Cartesian grid
\begin{equation}
    \mathcal{G}^{(k)}
    =
    \mathcal{D}_z^{(k)} \times \mathcal{D}_y^{(k)}
    =
    \left\{
        \left(
            B_{z,i}^{(k)},B_{y,j}^{(k)}
        \right)
    \right\},
\end{equation}
where $\mathcal{D}_z^{(k)}$ and $\mathcal{D}_y^{(k)}$ denote the current
one-dimensional supports of $B_z$ and $B_y$, respectively. The corresponding
posterior density values are
\begin{equation}
    p_{ij}^{(k)}
    =
    p\left(
        B_{z,i}^{(k)},B_{y,j}^{(k)}
        \mid
        \mathbf{y}_{1:k}
    \right).
\end{equation}

The initial prior is uniform over the chosen physical parameter region,
\begin{equation}
    p_0(\mathbf{B}) \propto 1,
    \qquad
    B_z\in\mathcal{D}_z^{(0)},
    \qquad
    B_y\in\mathcal{D}_y^{(0)},
\end{equation}
and vanishes outside it. For uniform grid spacings
$\Delta B_z^{(k)}$ and $\Delta B_y^{(k)}$, the discrete posterior is
normalized according to
\begin{equation}
    \sum_{i,j}
    p_{ij}^{(k)}
    \Delta B_z^{(k)}
    \Delta B_y^{(k)}
    =1.
\end{equation}

As successive measurements are incorporated, the posterior may become
localized on a scale comparable to the current grid spacing. Continuing the update on an insufficiently resolved grid can distort the posterior geometry and may cause the numerically represented posterior to vanish. To maintain
numerical resolution, the grid is refined when the posterior becomes
insufficiently resolved. The refinement region $\mathcal{R}^{(k)}$ is chosen
as the smallest rectangular domain containing a fraction $\alpha$ of the
current posterior mass,
\begin{equation}
    \int_{\mathcal{R}^{(k)}}
    p\left(
        \mathbf{B}\mid\mathbf{y}_{1:k}
    \right)
    d\mathbf{B}
    \geq \alpha,
\end{equation}
with $\alpha \simeq 0.999$ in the present implementation.

If refinement is triggered, the grid for the next update,
$\mathcal{G}^{(k+1)}$, is constructed with finer spacing over
$\mathcal{R}^{(k)}$. Otherwise,
$\mathcal{G}^{(k+1)}=\mathcal{G}^{(k)}$.

% Rather than interpolating the existing posterior onto a refined grid, all
% measurements incorporated up to step $k$ are replayed on the new grid:
Posterior on the new and refined grid is obtained by replaying and updating over all measurements incorporated up to step $k$:
\begin{equation}
    p\left(
        \mathbf{B}^{(k+1)}
        \mid
        \mathbf{y}_{1:k}
    \right)
    \propto
    p_0
    \left(
        \mathbf{B}^{(k+1)}
    \right)
    \prod_{\ell=1}^{k}
    \widetilde{\mathcal{L}}_{\ell}
    \left(
        \mathbf{B}^{(k+1)}
    \right),
    \label{eq:posterior_replay}
\end{equation}
where $\mathbf{B}^{(k+1)}$ denotes points on the refined grid and
$p_0$ is uniform over its support. 
% Each $\widetilde{\mathcal{L}}_{\ell}$ includes the bias-field marginalization of Eq.~\eqref{eq:bias_marginalized_likelihood}.

Thus, at each refinement, at most \(0.1\%\) of the represented posterior probability is discarded in order to reduce the computational domain while retaining essentially all of the posterior support. The posterior on the refined grid is then reconstructed from the complete measurement record.

\subsection{Adaptive Measurement Selection via Expected Information Gain}
\label{sec:expected_information_gain}

After each posterior update, the longitudinal bias field for the next
measurement is selected using the expected information gain. For a candidate
bias field \(B_c\) and a potential future observation \(\mathbf{y}\), the
information gained about the joint field vector is quantified by the
Kullback-Leibler divergence,
\begin{multline}
    D_{\mathrm{KL}}
    \left[
        p\left(
            \mathbf{B}
            \mid
            \mathbf{y},B_c,\mathbf{y}_{1:k}
        \right)
        \,\middle\|\,
        p_k(\mathbf{B})
    \right]
    = \\
    \int
    p\left(
        \mathbf{B}
        \mid
        \mathbf{y},B_c,\mathbf{y}_{1:k}
    \right)
    \log
    \frac{
        p\left(
            \mathbf{B}
            \mid
            \mathbf{y},B_c,\mathbf{y}_{1:k}
        \right)
    }{
        p_k(\mathbf{B})
    }
    d\mathbf{B},
    \label{eq:joint_kl_divergence}
\end{multline}
where, the $p_k\left( \mathbf{B}\right) \equiv p\left(
            \mathbf{B}
            \mid
            \mathbf{y}_{1:k}
        \right) $ is the current posterior.

Because the outcome of the next measurement is not known in advance, the
utility of a candidate bias field is obtained by averaging this information
gain over the predictive distribution, the probability of the hypothetical outcome $\mathbf{y}$ given the data so far:
\begin{multline}
    U_k(B_c)
    =
    \int
    p\left(
        \mathbf{y}
        \mid
        B_c,\mathbf{y}_{1:k}
    \right)
   \\ D_{\mathrm{KL}}
    \left[
        p\left(
            \mathbf{B}
            \mid
            \mathbf{y},B_c,\mathbf{y}_{1:k}
        \right)
        \,\middle\|\,
        p_k(\mathbf{B})
    \right]
    d\mathbf{y}.
    \label{eq:expected_information_gain}
\end{multline}
The predictive distribution is
\begin{equation}
    p\left(
        \mathbf{y}
        \mid
        B_c,\mathbf{y}_{1:k}
    \right)
    =
    \int
    p\left(
        \mathbf{y}
        \mid
        \mathbf{B},B_c
    \right)
    p_k(\mathbf{B})
    d\mathbf{B}.
    \label{eq:posterior_predictive}
\end{equation}
The next requested bias-field setting is selected as
\begin{equation}
    B_{c,*}^{(k+1)}
    =
    \underset{B_c\in\mathcal{B}_c}{\operatorname{arg\,max}}
    \;
    U_k(B_c),
    \label{eq:optimal_bias_selection}
\end{equation}
where \(\mathcal{B}_c\) is the set of available bias-field set-points. 
Direct numerical evaluation of these nested integrals,
\begin{multline}
    U_k(B_c)
    =
    \int d\mathbf{y}
    \int d\mathbf{B^{'}}\,
    p\left(
        \mathbf{y}
        \mid
        \mathbf{B^{'}},B_c
    \right)
    p_k(\mathbf{B^{'}})
    \times\\
    \left[
        \int d\mathbf{B^{''}}\,
        p\left(
            \mathbf{B^{''}}
            \mid
            \mathbf{y},B_c,\mathbf{y}_{1:k}
        \right)
        \log
        \frac{
            p\left(
                \mathbf{B^{''}}
                \mid
                \mathbf{y},B_c,\mathbf{y}_{1:k}
            \right)
        }{
            p_k(\mathbf{B^{''}})
        }
    \right],
    \label{eq:eig_expanded}
\end{multline} 

 is computationally
expensive. Therefore the integrals are approximated using Monte Carlo sampling. 
$N_{\rm MC}$ field samples from the current joint posterior are drawn,
\begin{equation}
    \mathbf{B}_i
    \sim
p_k(\mathbf{B})
    \qquad
    i=1,\ldots,N_{\rm MC},
\end{equation}
and, for each sampled field and candidate bias \(B_c\), a
hypothetical observation
\begin{equation}
    \mathbf{y}_i
    \sim
    p\left(
        \mathbf{y}
        \mid
        \mathbf{B}_i,B_c
    \right)
\end{equation}
is sampled.
The pairs
\(\left(\mathbf{B}_i,\mathbf{y}_i\right)\)
are therefore sampled from the joint predictive distribution
\begin{equation}
    p \left( \textbf{B}, \textbf{y} \mid B_c\right) 
    =
    p_k(\mathbf{B}) \,
    p(\mathbf{y}
        \mid
        \mathbf{B},B_c).
\end{equation}
The outer expectation in Eq.~\eqref{eq:eig_expanded} is then approximated as
\begin{multline}
    U_k(B_c)
    \approx
    \frac{1}{N_{\rm MC}}
    \sum_{i=1}^{N_{\rm MC}}
    D_{\rm KL}
    \!\left[
    p(\mathbf B|\mathbf y_i,B_c,\mathbf y_{1:k})
    \,\|\,p_k(\mathbf B)
    \right].
        \label{eq:eig_mc_outer}
\end{multline}
For each hypothetical outcome
\(\mathbf{y}_i\), Bayes' rule gives
\begin{equation}
    p\left(
        \mathbf{B}
        \mid
        \mathbf{y}_i,B_c,\mathbf{y}_{1:k}
    \right)
    =
    \frac{
        p\left(
            \mathbf{y}_i
            \mid
            \mathbf{B},B_c
        \right)
        p_k(\mathbf{B})
    }{
        \displaystyle
        \int
        p\left(
            \mathbf{y}_i
            \mid
            \mathbf{B},B_c
        \right)
        p_k(\mathbf{B})
        d\mathbf{B}
    }.
    \label{eq:hypothetical_posterior_mc}
\end{equation}
and the remaining integral over \(\mathbf B\) is evaluated deterministically on the current posterior grid \(G^{(k)}\) introduced in Sec.~\ref{sec:posterior_grid_refinement}.
The normalization in the denominator is approximated by
\begin{multline}
    p(\mathbf y_i|B_c,\mathbf y_{1:k}) \\
    \approx
    \sum_{r,s}
    p(\mathbf y_i|
    B_{z,r}^{(k)},B_{y,s}^{(k)},B_c)
    \,
    p_{rs}^{(k)}
    \Delta B_z^{(k)}
    \Delta B_y^{(k)} .
\end{multline}
The posterior density on the same grid after the hypothetical observation \(\mathbf y_i\) is

\begin{gather}
    p_{rs}^{(k,i)}(B_c) =
    \frac{
    p(\mathbf y_i|
    B_{z,r}^{(k)},B_{y,s}^{(k)},B_c)
    \, p_{rs}^{(k)}
    }{
    p(\mathbf y_i|B_c,\mathbf y_{1:k})
    }.
\end{gather}
The corresponding KL divergence is evaluated on the grid as
\begin{multline}
D_{\rm KL}^{(i)}(B_c)\\
\approx
\sum_{r,s}
p_{rs}^{(k,i)}(B_c)
\log
\frac{
p_{rs}^{(k,i)}(B_c)
}{
p_{rs}^{(k)}
}
\Delta B_z^{(k)}
\Delta B_y^{(k)} .
\end{multline}
The expected information gain is then estimated as
\begin{gather}
U_k(B_c)
\approx
\frac{1}{N_{\rm MC}}
\sum_{i=1}^{N_{\rm MC}}
D_{\rm KL}^{(i)}(B_c).
\label{eq:eig_mc_sum}
\end{gather}

Equation~\eqref{eq:eig_mc_sum} approximates the expected KL divergence
by averaging over $\mathrm{N}_{\mathrm{MC}}$ possible future observations. \(N_{\mathrm{MC}}=400\) Monte Carlo samples per candidate bias field, was chosen to keep the repeated utility evaluations computationally tractable. The same set of posterior samples is reused for all candidate bias fields to sample possible observations.
In addition, although the Bayesian update incorporates uncertainty in the applied bias field through the marginal likelihood described in Sec.~\ref{sec:joint_bayesian_update}, this bias field marginalization when evaluating the expected information gain is not repeated. Instead, the utility for a candidate setting \(B_c^\ast\) is evaluated using the nominal candidate value \(B_c=B_c^\ast\). This approximation substantially reduces the computational cost of the adaptive design calculation. Consequently, the expected information gain used for bias selection should be regarded as an approximate design utility, whereas the subsequent posterior update retains the treatment of bias-field uncertainty described above
% In the present implementation, a complete \(0.2~\mu\mathrm{s}\) temporal response contains several experimental sampling points. 
Evaluating the hypothetical likelihood over every time
sample for every posterior draw and every candidate bias field is
computationally expensive. Therefore the expected utility is evaluated on a
sub-sampled temporal grid, for every fourth time sample of the next measurement block.

% The bias uncertainty of
% Eq.~\eqref{eq:bias_uncertainty_prior} is also included in the utility
% calculation. Consequently, the hypothetical likelihood associated with a
% requested candidate \(B_c\) is replaced by its bias-marginalized form,
% %
% \begin{equation}
%     \widetilde{p}
%     \left(
%         \mathbf{y}
%         \mid
%         \mathbf{B},B_c
%     \right)
%     =
%     \int
%     p\left(
%         \mathbf{y}
%         \mid
%         \mathbf{B},B_c'
%     \right)
%     \pi_c
%     \left(
%         B_c'
%         \mid
%         B_c
%     \right)
%     dB_c'.
%     \label{eq:eig_bias_marginalization}
% \end{equation}
% %
% Thus, the selected bias maximizes the information expected from the
% experiment after accounting for uncertainty in the magnetic-field
% parameters and uncertainty in the realized control field.
Thus, the selected bias field maximizes the information expected from the next
measurement, averaged over the possible magnetic field configurations represented by
the current joint posterior.

% TODO: State N_MC and whether posterior samples are drawn with replacement.
%
% TODO: State the number of candidate bias fields and their numerical range.
%
% TODO: State whether the same posterior and noise samples are reused for
% every candidate B_c. Reusing common random numbers reduces Monte Carlo
% fluctuations in comparisons between candidate utilities.
%
% TODO FIGURE: Plot U_k(B_c) at several representative update times and mark
% the selected maximum.
%
% TODO FIGURE: Plot the selected B_c trajectory over the experimental signal
% map, alongside the final MAP estimate.

\subsection{Offline Adaptive-search Implementation}
\label{sec:offline_implementation}

The present implementation evaluates the joint Bayesian inference and
adaptive measurement selection policy in post-processing using a
pre-recorded experimental dataset. 
It should therefore be distinguished from a real-time feedback controller.

In the intended experimental protocol, the $k^{\mathrm{th}}$ Bayesian update would use an independent measurement realization, which would consist of preparing and optically pumping a fresh atomic ensemble, applying a fixed bias field \(B_c^{(k)}\), and recording the complete transient response.
The resulting observation would be used to update the joint posterior and select the bias field applied during the next independently prepared realization:
\begin{multline}
    \text{Prepare}
    \rightarrow
    \text{Pump}
    \rightarrow
    \text{Apply } B_c^{(k)}
    \rightarrow 
    \text{Record transient} \\
    \rightarrow
    \text{Update posterior}
    \rightarrow
    \text{Select } B_c^{(k+1)}
    \rightarrow
    \text{Repeat}.
    \label{eq:online_implementation}
\end{multline}

In the present offline implementation, the adaptive policy does not control the experiment during acquisition. The archived dataset described in Sec.~\ref{sec:setup_cold_atom_platform} does not contain a sequence of full transient
realizations acquired under bias fields chosen online by the algorithm. Instead, it queries the archived response surface. For the offline replay, the recorded transient is divided into sequential temporal intervals of duration \(\Delta t_{\rm step}=0.2~\mu{\rm s}\). The data ingestion for all datasets used were started at $t_0 = 1 \mu$s. The start time of the \(k^{\rm th}\) interval is denoted by
\[ 
t_k= t_0 + (k-1)\Delta t_{\rm step}, 
\]
such that the data used at update \(k\) correspond to \(t\in[t_k,t_k+\Delta t_{\rm step})\). At update \(k\), the algorithm selects \(B_c^{(k)}\) using the expected information gain criterion and retrieves the experimental samples in the \(k^{\rm th}\) temporal interval from the recorded transient corresponding to that bias. Therefore, $\mathbf{y}_{\mathrm{exp}}^{(k)} \equiv \{I_{\mathrm{exp}}(t,B_c^{(k)})\}$ with \(t\in[t_k,t_k+\Delta t_{\rm step})\). After updating the posterior, the algorithm selects \(B_c^{(k+1)}\), advances to the subsequent temporal interval, and queries the corresponding independently recorded transient.
This construction allows the posterior evolution and the expected information gain policy to be examined using real experimental observations without yet implementing the required real-time experimental control.

The resulting bias-field trajectory should therefore be interpreted as the sequence of measurement settings requested as an offline adaptive search, rather than as a bias field that was physically changed during the evolution of a single atomic ensemble. 
In particular, the forward model used in every likelihood evaluation assumes that the bias field remains constant over the corresponding simulated transient.

To emulate finite uncertainty in the applied control field during the offline replay, each selected nominal bias \(B_c^{(k)}\) is associated with a uniform uncertainty interval \(B_c^{(k)}\pm\delta B_c\), where \(\delta B_c\) is chosen from the resolution of the available experimental bias settings. The likelihood used for the Bayesian update is then marginalized over this interval as described in Sec.~\ref{sec:joint_bayesian_update}. This procedure does not represent a separately measured uncertainty of the experimental bias field. Rather, it provides an offline model of finite control uncertainty. Because the assumed distribution is symmetric about the nominal bias value, it introduces no directional bias in \(B_c\).

The offline analysis consequently tests whether the measured transient CPT response and the available forward model theory, together with joint Bayesian field estimation and information-guided selection of the longitudinal bias field is able to resolve the field components.
A direct online implementation would preserve the same inference and
measurement-selection procedure while acquiring a new complete atomic
transient at each requested bias setting.

To summarize, our offline evaluation proceeds as follows: 
\begin{multline*}
\begin{gathered}
\text{Turning on of Probe Pulse} \, \rightarrow \text{initialization of prior} \rightarrow  \\ \text{Observation of data at } B_c^{(k)} \rightarrow  \text{Bayes Update on the data} \\ \text{ in } [t_k,t_k+\Delta t_{\rm step})  \rightarrow \text{Select $B_c^{(k+1)}$}  \rightarrow \\ \text{Repeat till end of scan.} 
\end{gathered}
\end{multline*}

\section{Results}
\label{sec:results}

Unless otherwise stated, the adaptive-inference results presented below use a likelihood scale of \[\sigma = 0.0045 \, I_{\mathrm{sat}} .\]

The dependence of the inferred fields on this choice and its relation to the independently characterized experimental noise, are examined in Sec.~\ref{sec:sigma_dependence}.

\subsection{Posterior Convergence and Uncertainty Contraction}
\label{sec:estimate_convergence}

\begin{figure*}[ht]
    \centering
    \includegraphics[width=\linewidth]{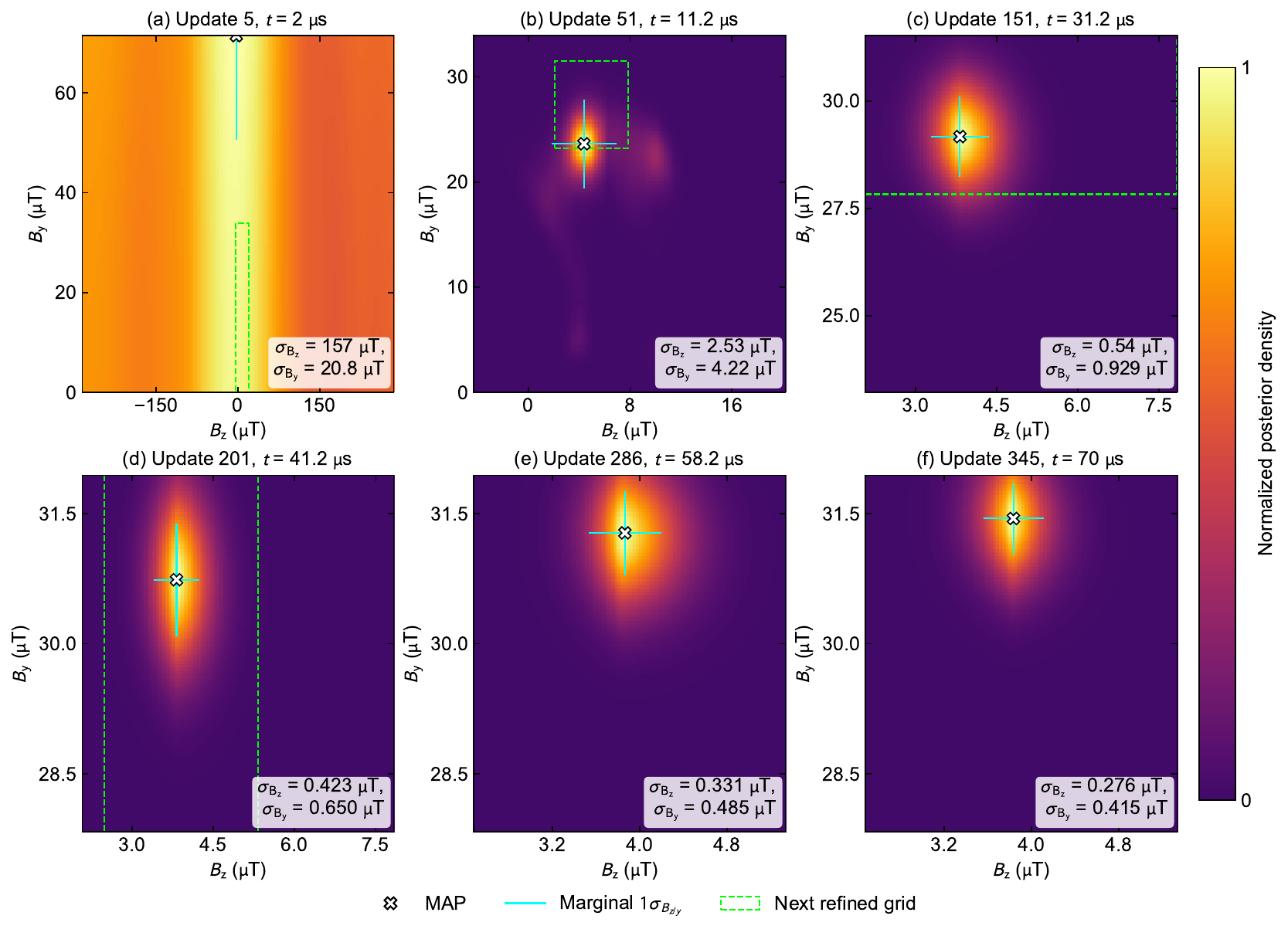}
    \caption{Evolution of the joint posterior $p(B_z,B_y\mid\mathbf{y})$ for the representative $B_{y,\mathrm{FID}}=35~\mu\mathrm{T}$ dataset. The cross marks the MAP estimate, cyan bars show the marginal posterior standard deviations, and dashed rectangles indicate the region retained for subsequent grid refinement. The displayed parameter ranges change as the posterior grid is refined.}
    \label{fig:posterior_evolution}
\end{figure*}

The convergence of the joint posterior
$p \left( \mathbf{B}\mid\mathbf{y}\right)$, with $\mathbf{B}=(B_z,B_y)^{\mathsf T}$, is first examined for a representative transverse field scan. Figure~\ref{fig:posterior_evolution} shows selected snapshots of the posterior during the offline adaptive-search sequence. Starting from the broad initial prior, the distribution rapidly contracts in $B_z$ and $B_y$, until a compact joint posterior is obtained at the end of the recorded transient. The changing physical ranges of the axes reflect the adaptive refinement of the posterior support grid over $B_z$ and $B_y$. The dashed rectangles indicate the high-probability regions retained for subsequent refinement which occurs in the interim, while the MAP estimate and the marginal posterior standard deviations show the evolution of the inferred field components and their uncertainties.

\begin{figure}[ht]
    \centering
    \includegraphics[width=\linewidth]{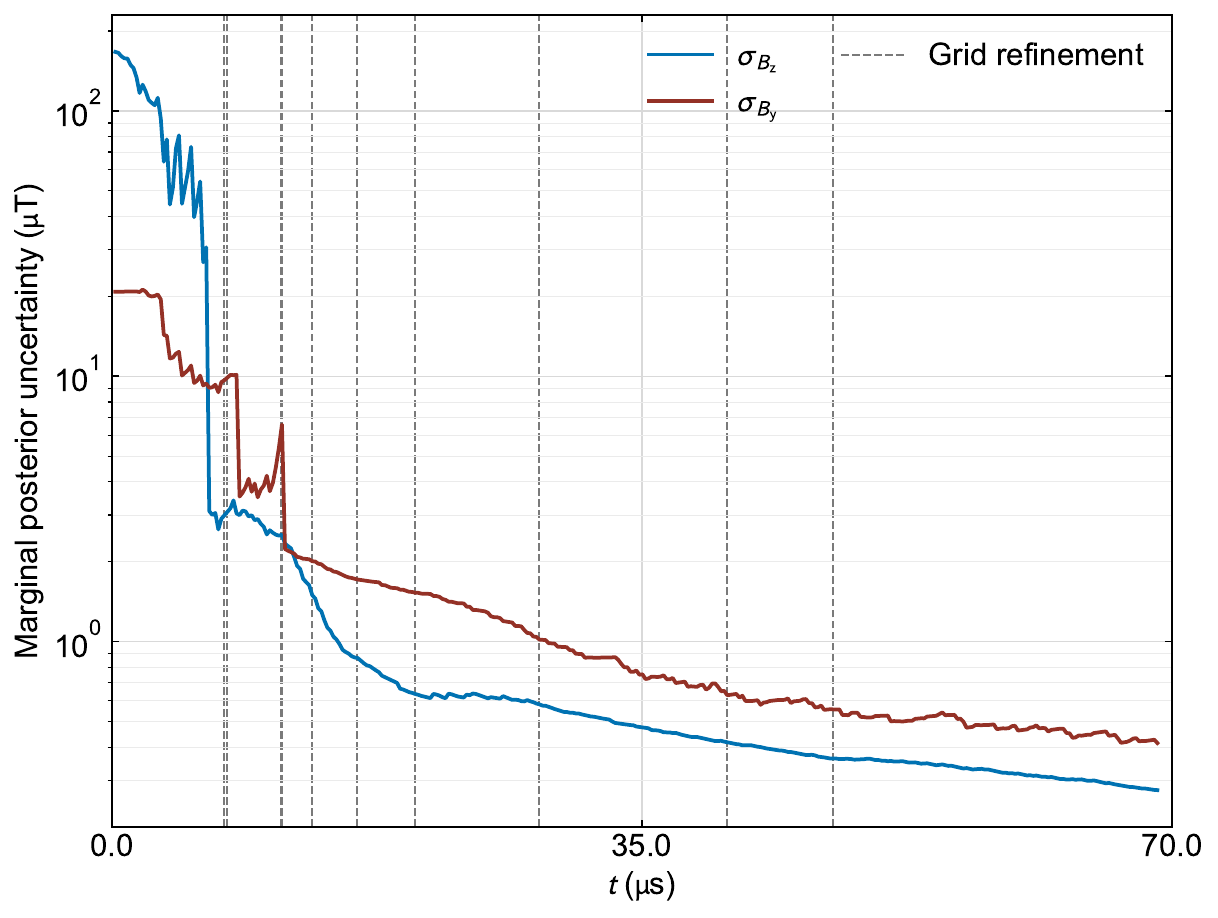}
    \caption{
    Marginal posterior uncertainties $\sigma_{B_z}$ and $\sigma_{B_y}$ during the  representative $B_{y,\mathrm{FID}}=35~\mu\mathrm{T}$ adaptive-search run. Vertical dashed lines mark grid-refinement events.}
    \label{fig:uncertainty_evolution}
\end{figure}

The contraction of the marginal uncertainties is shown more directly in Fig.~\ref{fig:uncertainty_evolution}. The quantities
$\sigma_{B_z}=\sqrt{\operatorname{Var}(B_z)}$ and
$\sigma_{B_y}=\sqrt{\operatorname{Var}(B_y)}$
decrease as the posterior is updated over successive temporal data blocks. The vertical dashed lines in Fig.~\ref{fig:uncertainty_evolution} indicate grid refinement events. Refinement is triggered by insufficient numerical resolution of the posterior, ensuring that both gradual and rapid posterior contraction remain adequately resolved. The continued decrease in $\sigma_{B_z}$ and $\sigma_{B_y}$ reflects the progressive narrowing of the posterior as the knowledge about the field is updated on successive measurements. These widths describe the posterior under the adopted forward model and likelihood scale $\sigma$; they are
not calibrated credible intervals for the physical field.

For the representative scan shown in Figs.~\ref{fig:posterior_evolution} and~\ref{fig:uncertainty_evolution}, the final joint maximum-a-posteriori MAP estimate is 
\begin{equation}
\widehat{\mathbf{B}}_{\mathrm{MAP}}
=
\begin{pmatrix}
3.83 \\
31.44
\end{pmatrix}
\mu\mathrm{T},
\end{equation}
with posterior covariance matrix
\begin{equation}
\boldsymbol{\Sigma}_{\mathbf{B}}
=
\begin{pmatrix}
0.0760 & -0.0051 \\
-0.0051 & 0.1720
\end{pmatrix}
\left(\mu\mathrm{T}\right)^{2}.
\end{equation}
The corresponding marginal posterior standard deviations are
\begin{equation}
\sigma_{B_z}\approx0.28~\mu\mathrm{T},
\qquad
\sigma_{B_y}\approx0.41~\mu\mathrm{T},
\end{equation}
and the posterior correlation coefficient is
\begin{equation}
\rho_{zy}
=
\frac{\operatorname{Cov}(B_z,B_y)}
{\sqrt{\operatorname{Var}(B_z)\operatorname{Var}(B_y)}}
\approx -0.045.
\end{equation}

 The posterior standard deviation represents our uncertainty in the field under the adopted forward model and likelihood scale $\sigma$ only.
The small magnitude of $\rho_{zy}$ indicates only weak linear correlation between the inferred longitudinal and transverse field components for this dataset. This is consistent with their distinct effects on the modeled CPT response: variations in $B_z$ predominantly shift the longitudinal CPT resonance peak, whereas $B_y$ modifies the CPT linewidth and lineshape through transverse-state mixing. The weak final covariance therefore provides evidence that these two field components remain distinguishable within the adopted forward model and the measured transient response. Whether the physical components are separately recovered is examined
against independent calibration in Sec.~\ref{sec:transverse_field_consistency}.

\subsection{Adaptive Bias Selection Behavior}
\label{sec:adaptive_bias_selection}

\begin{figure}[ht]
    \centering
    \includegraphics[width=\linewidth]{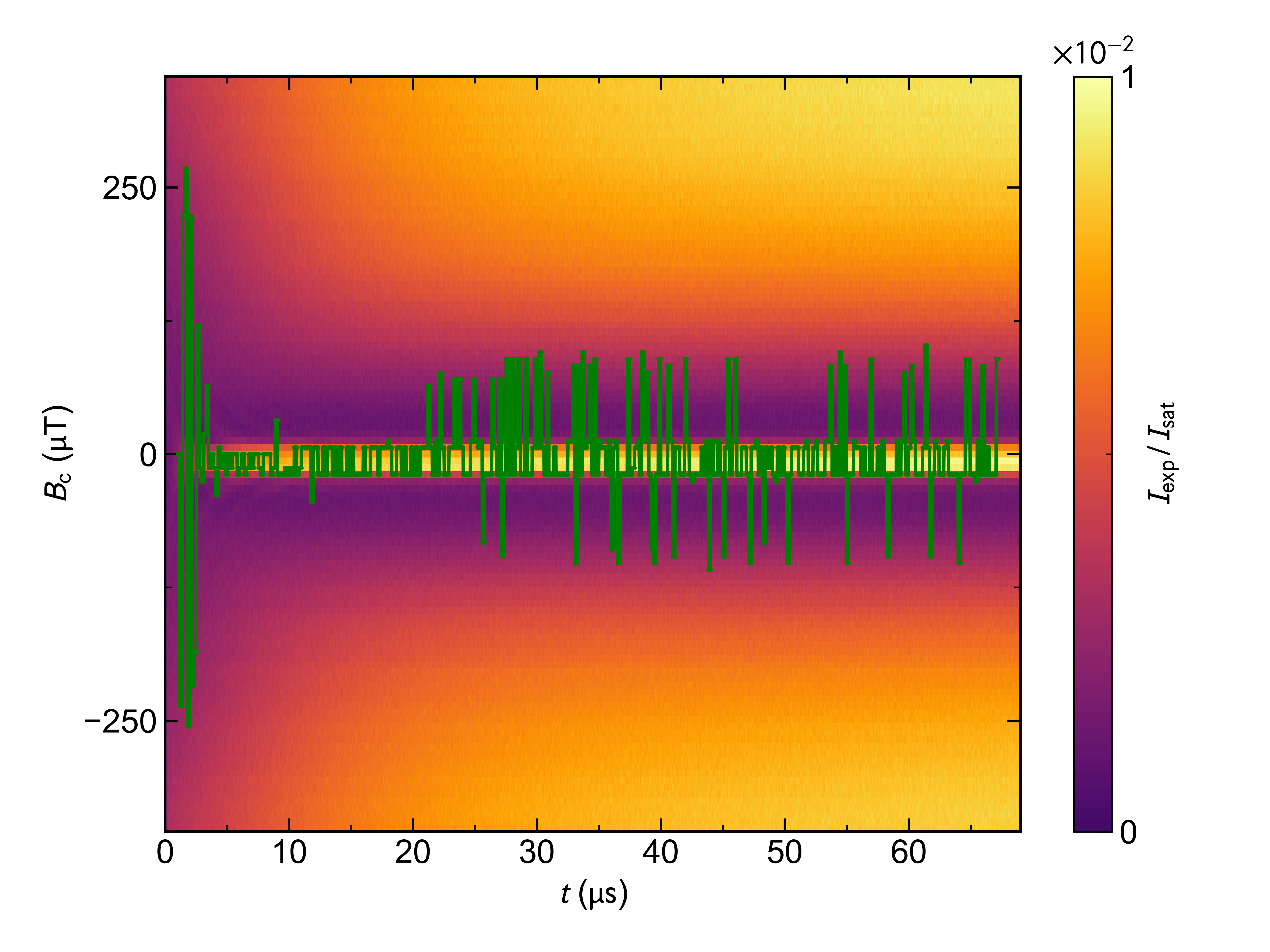}
    \caption{Adaptive longitudinal bias-field selection for the representative $B_{y,\mathrm{FID}}=35~\mu\mathrm{T}$ dataset. The green trajectory shows the requested bias field $B_c^{(k)}$, overlaid on the measured probe transmission $I_{\mathrm{exp}}/I_{\mathrm{sat}}$. The trajectory denotes the sequence queried by the offline adaptive search.}
    \label{fig:bias_trajectory}
\end{figure}

Figure~\ref{fig:bias_trajectory} shows the longitudinal bias fields $B_c^{(k)}$ selected by the expected information gain policy during the run. At early times, when the joint posterior $p(\mathbf{B}\mid\mathbf{y}_{1:k})$ remains broad, the selected settings span a comparatively wide range of available bias fields. As the posterior contracts, the search becomes increasingly concentrated within a
narrower region of $B_c$, while retaining occasional choices of more distant measurement settings. The selected trajectory therefore evolves together with the inferred joint posterior rather than following a predetermined scan sequence, as is expected.

The selected bias values should not, at present, be interpreted simply as points of maximum local slope of the CPT response. The utility $U_k(B_c)$ is evaluated with respect to the full joint posterior over $\mathbf{B}=(B_z,B_y)^{\mathsf T}$, and therefore quantifies the expected information gained about both field components simultaneously. A measurement configuration that is highly sensitive to one component alone need not maximize the expected reduction in uncertainty of the joint field estimate. Although the selected settings frequently lie in regions where the optical response varies appreciably with the applied longitudinal bias, establishing
a quantitative connection between the finite-posterior expected-information-
gain criterion and local gradient sensitivity requires more work and a separate analysis.

As stated earlier, within the present post-processing implementation, the trajectory in Fig.~\ref{fig:bias_trajectory} represents the sequence of bias settings requested by the adaptive search. It should not be interpreted as a bias field that was physically varied during the evolution of a single atomic
ensemble. 

\subsection{Consistency across Applied Transverse Fields}
\label{sec:transverse_field_consistency}

To examine the consistency of the joint Bayesian estimator across different magnetic-field configurations, the offline procedure is repeated for four experimental scans in which the applied transverse field is varied while the ambient longitudinal field is nominally held fixed. The applied transverse field, denoted ($B_{y,\mathrm{FID}}$), is independently calibrated using a FID measurement performed after compensating the residual $B_x$ and $B_z$ components.
The $B_{y,\mathrm{FID}}=35~\mu\mathrm{T}$ dataset is the representative case used in the preceding sections to illustrate the posterior evolution, uncertainty contraction, and adaptive bias-selection behavior.

\begin{figure}[ht]
    \centering
    \includegraphics[width=\columnwidth]{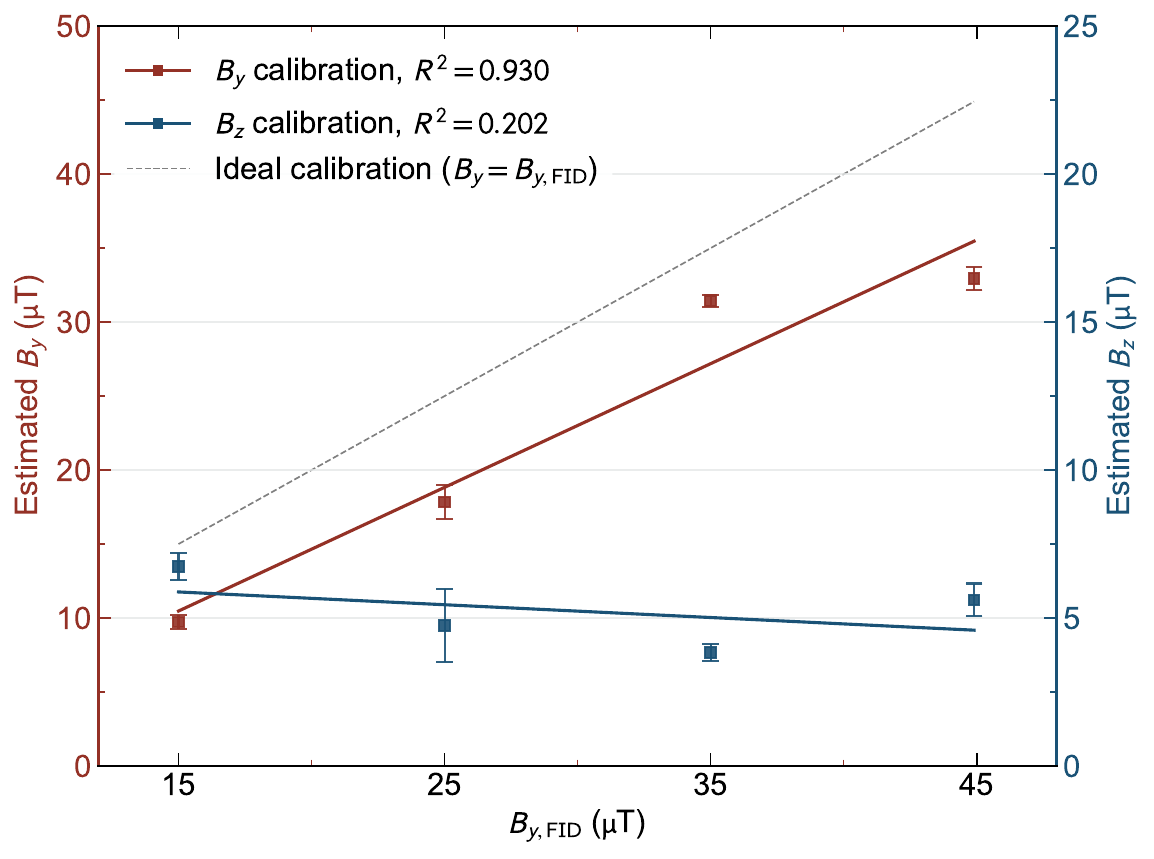}
    \caption{Calibration of the final joint MAP estimates across the applied transverse field datasets. $\hat B_y$ and $\hat B_z$ are shown as functions of the independently calibrated $B_{y,\mathrm{FID}}$, with linear fits indicated by solid lines. The dashed line denotes the ideal transverse field relation $\hat B_y=B_{y,\mathrm{FID}}$. Error bars are the marginal posterior standard deviations and represent precision under the adopted forward model only.}
    \label{fig:calibration}
\end{figure}

Figure~\ref{fig:calibration} shows the final joint MAP estimates obtained for each value of \(B_{y,\mathrm{FID}}\). The inferred transverse component follows the applied field approximately linearly,
\begin{equation}
    \widehat{B}_{y,\mathrm{MAP}}
    =
    (0.836 \pm 0.162)\,
    B_{y,\mathrm{FID}}
    -
    2.068~\mu\mathrm{T},
    \label{eq:By_calibration}
\end{equation}
with a coefficient of determination ($R^2=0.930$). The estimator therefore tracks the variation of the transverse field across the investigated range, although the fitted slope differs appreciably from unity. This discrepancy is discussed in Sec.~\ref{sec:forward_model_fidelity},
where residual differences between the experimental signal and the forward model  are examined and the possible role of unmodeled nuisance parameters is discussed.

The inferred longitudinal component exhibits a substantially weaker dependence
on the applied transverse field sweep. A linear fit gives
\begin{equation}
    \widehat{B}_{z,\mathrm{MAP}}
    =
    (-0.043 \pm 0.061)\,
    B_{y,\mathrm{FID}}
    +
    6.528~\mu\mathrm{T},
    \label{eq:Bz_calibration}
\end{equation}
with ($R^2=0.202$). Since $B_z$ is nominally unchanged between these measurements, the relevant diagnostic is primarily the fitted slope. The slope is small compared with the response of the inferred $B_y$ component, indicating limited cross-axis dependence over the measured range.

Taken together, these results show that the joint estimator distinguishes the different roles of the two magnetic-field components in the measured CPT response: changes in the independently applied transverse field are recovered
predominantly through $B_y$, while the inferred $B_z$ remains comparatively stable. The remaining systematic deviation of \(\widehat{B}_{y,\mathrm{MAP}}\) from the independently calibrated field, however, shows that posterior convergence alone does not establish absolute accuracy and motivates the forward-model analysis discussed below.

\subsection{Dependence on the Likelihood Scale $\sigma$}
\label{sec:sigma_dependence}

Because the likelihood scale controls how strongly model-data
differences influence the posterior, the dependence of the inferred
fields on $\sigma$ is examined before interpreting the results obtained at
the value used in the preceding sections.

The experimental and simulated probe intensities are expressed in units of $I_{\mathrm{sat}}$. The experimentally
characterized intensity fluctuations are therefore directly comparable with
the confidence scale $\sigma$ used in the Gaussian likelihood.

\begin{figure}[htbp]

    \centering
    \includegraphics[width=\columnwidth]
    {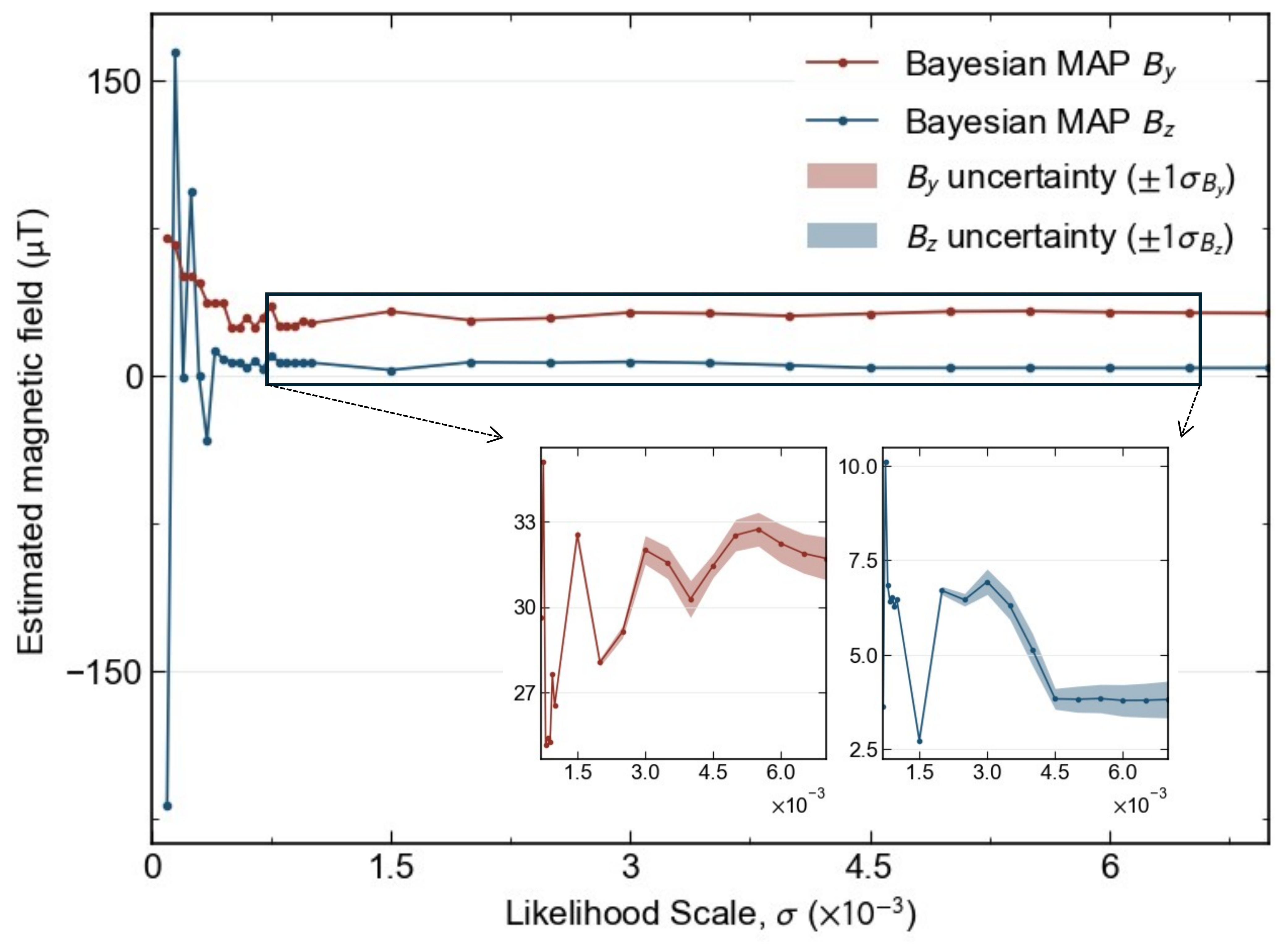}

    \caption{Dependence of the final joint MAP estimates on the likelihood scale $\sigma$ for the representative $B_{y,\mathrm{FID}}=35~\mu\mathrm{T}$ dataset. Shaded regions indicate the marginal posterior standard deviations. The insets highlight how the inferred $B_y$ and $B_z$ estimates progressively approach comparatively stable values as $\sigma$ is increased.}
    \label{fig:dependence_of_estimates_on_sigma}
\end{figure}

Figure~\ref{fig:dependence_of_estimates_on_sigma} shows the final joint MAP
estimates obtained over a range of likelihood scales for the representative $B_{y,\mathrm{FID}}=35~\mu\mathrm{T}$ dataset. Similar trends are observed for the remaining datasets. At the smallest values of $\sigma$, the inferred
fields can vary strongly and the corresponding posterior uncertainties can
become very small. As $\sigma$ is increased, the estimates generally enter
broader regions in which the dominant MAP values are substantially less
sensitive to further changes in the likelihood scale, while the posterior
uncertainties increase as expected from the broader likelihood.

Importantly, the likelihood scales over which stable estimates are obtained
are larger than the independently characterized experimental intensity noise ($\sim 10^{-4} \, I_{\mathrm{sat}}$, (Sec.~\ref{sec:measurement_model})).
The required likelihood width therefore cannot be identified with detector
noise alone. Rather, it acts as an effective confidence scale that also
accommodates disagreement between the experimental response and the present
forward model.

The results do not identify a unique optimal value of $\sigma$. Instead, they
show a range over which the inferred fields are comparatively stable, followed
at sufficiently large $\sigma$ by increasing posterior uncertainty as the
likelihood becomes less discriminating.

\subsection{Comparison with a Full-Scan Hypothesis Test}
\label{sec:full_scan_test}

As an additional consistency check, the complete experimental
$(t,B_c)$ scan is compared with the discrete $(B_z,B_y)$ hypotheses available on the precomputed simulation grid. 
This was not a continuous Bayesian inference; the
likelihood was evaluated directly over the fixed grid and the maximizing
hypothesis was identified.

The full-scan and adaptive estimates remain broadly compatible in $B_z$, but
differ substantially in $B_y$: for several datasets, the full scan $B_y$ estimates remain clearly separated from the corresponding adaptive posterior regions (Table~\ref{tab:estimate_comparison}). 
The full-scan hypotheses also produce
smaller global residuals over the complete measured response. However, their
$B_y$ calibration remains systematically different from the independently
calibrated field, with a fitted slope substantially different from unity.
Therefore, the lower global residual cannot be interpreted as evidence of
greater field-estimation accuracy.

Rather, the two procedures favor different compromises within the present
forward model: the adaptive estimator is constrained primarily by the
information-selected subset of the scan, whereas the full-scan test weights
the complete measured response. The corresponding calibration and residual
structure are shown in Appendix~\ref{app:full_scan_test_appendix}.

Across both the adaptive and full-scan comparisons, the magnitude of the
remaining model-data residuals tends to increase for the datasets acquired at
larger transverse fields. The origin of this trend is not established here and
may involve field-dependent sensitivity to presently fixed or omitted
forward-model parameters, which is discussed in Sec.~\ref{sec:forward_model_fidelity}.

\begin{table*}[htb]
    \caption{Comparison of the adaptive Bayesian estimates with the maximizing full-scan hypotheses. Bayesian uncertainties are marginal posterior standard deviations. The full-scan values $\hat{B}^*_z$ and $\hat{B}^*_y$ are evaluated on a discrete grid with spacings $\Delta B_z^{\mathrm{grid}}=7.66~\mu\mathrm{T}$ and $\Delta B_y^{\mathrm{grid}}=1.43~\mu\mathrm{T}$, which set the numerical resolution of the hypothesis test and are not statistical credible intervals.}
    \label{tab:estimate_comparison}
    \centering
    % Shrink the space between columns just enough to fit a standard column
    \setlength{\tabcolsep}{4pt}
    \begin{tabular}{lcccc}
        \toprule
        $B_{y, \mathrm{FID}}$ & \multicolumn{2}{c}{Bayesian MAP} & \multicolumn{2}{c}{Full-Scan Hypothesis Test} \\
        \cmidrule(lr){2-3} \cmidrule(lr){4-5} 
        ($\mathrm{\mu}$T) & $\hat{B}_z \, \pm \sigma_{B_z}$ & $\hat{B}_y \, \pm \, \sigma_{B_y}$ & $\hat{B}^*_z \, \pm \, \Delta B^{\mathrm{grid}}_{z}$ & $\hat{B}^*_y \, \pm \, \Delta B^{\mathrm{grid}}_{y}$ \\
        \midrule
        15.0 & $6.74 \pm 0.46$ & $9.72 \pm 0.47$ & $7.66 \pm 7.66$ & $17.14 \pm 1.43$ \\
        25.0 & $4.75 \pm 1.24$ & $17.84 \pm 1.16$ & $7.66 \pm 7.66$ & $32.86 \pm 1.43$ \\
        35.0 & $3.83 \pm 0.28$ & $31.44 \pm 0.41$ & $7.66 \pm 7.66$ & $47.14 \pm 1.43$ \\
        44.9 & $5.61 \pm 0.55$ & $32.95 \pm 0.79$ & $7.66 \pm 7.66$ & $60.00 \pm 1.43$ \\
        \bottomrule
    \end{tabular}
\end{table*}

\section{Discussion}
\label{sec:discussion}

\subsection{Forward Model Fidelity and Estimation Error}
\label{sec:forward_model_fidelity}

Although the joint Bayesian posterior converges to a well-localized estimate for each experimental scan, the calibration results of Sec.~\ref{sec:transverse_field_consistency} show that posterior concentration does not by itself establish absolute accuracy. In particular, the inferred transverse field tracks the independently calibrated $B_{y,\mathrm{FID}}$ approximately linearly, but with a slope that differs appreciably from unity. This combination of a narrow posterior and a systematically displaced MAP estimate is indicative of model discrepancy: the Bayesian update can identify the parameter values that best reproduce the experimental signal within the adopted forward model even when that model does not reproduce the underlying experiment exactly.

\begin{figure}[htbp]
    \centering
    \includegraphics[width=\columnwidth]{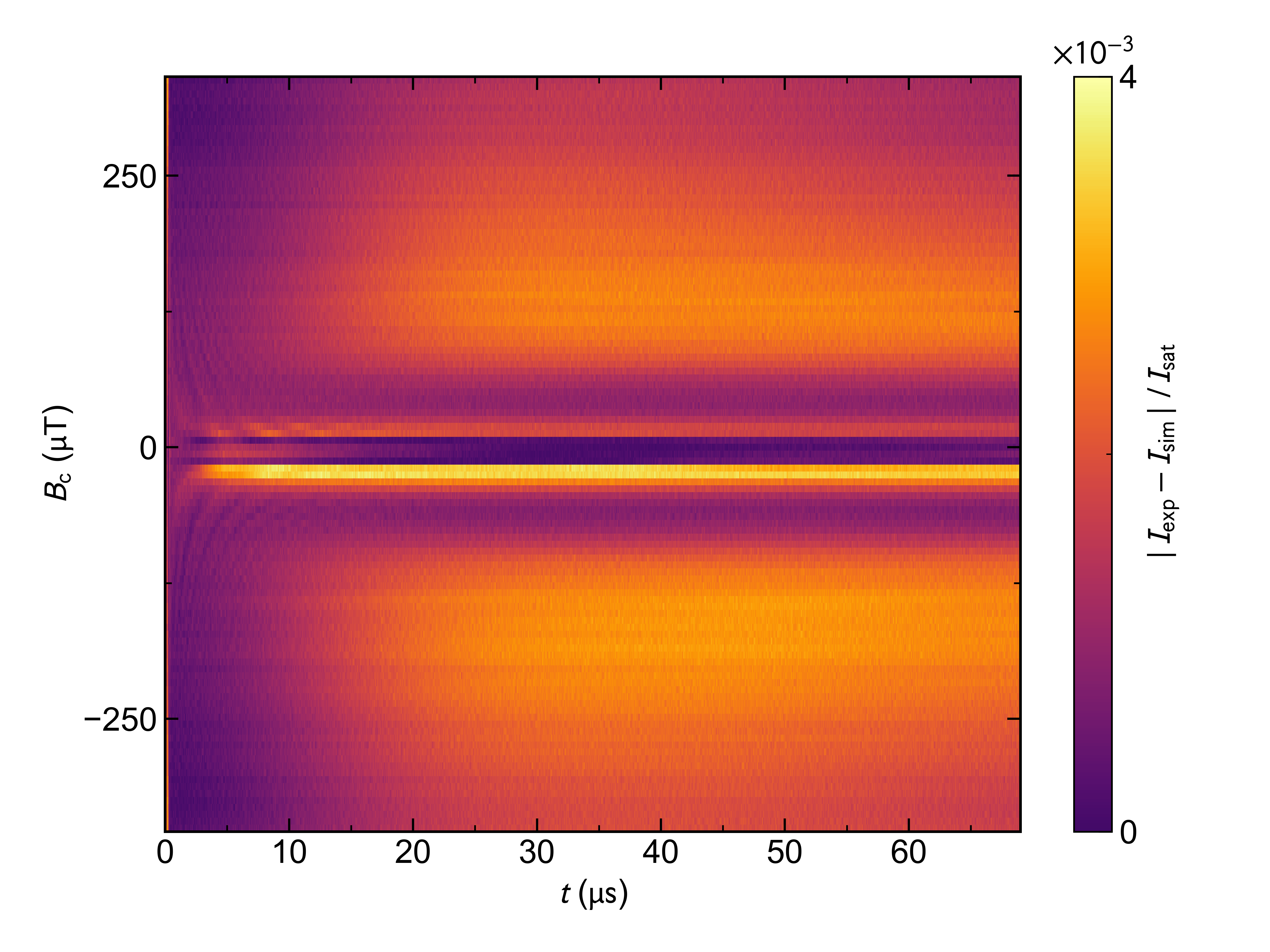}
        % \caption{$B_{y,\mathrm{FID}}$ Scan 2, Loss $= 1.3351449382037752$}

    \caption{Absolute model-data residual $R(t,B_c)$ evaluated at the final adaptive Bayesian estimate for the representative $B_{y,\mathrm{FID}}=35~\mu\mathrm{T}$ dataset. The summed squared residual over the displayed $(t,B_c)$ domain is $L=1.33 \, I^2_{\mathrm{sat}}$.}
    \label{fig:difference_plots}
\end{figure}

The structure of this discrepancy is visible more directly in the residual map,
\begin{multline}
    R(t,B_c) 
    = \\
    \left|
        I_{\mathrm{exp}}(t,B_c)
        -
        I_{\mathrm{sim}}
        \left(
            t,B_c;
            \widehat{B}_{z,\mathrm{MAP}},
            \widehat{B}_{y,\mathrm{MAP}}
        \right)
    \right|,
\end{multline}
shown in Fig.~\ref{fig:difference_plots}. For each transverse field dataset, the experimental and simulated CPT resonance positions agree well: the residual near the center of the CPT feature is low, indicating that the
inferred \(B_z\) reproduces the longitudinal location of the resonance at least to the correct scale. This is consistent with the role of \(B_z\) in the forward model, where the longitudinal field primarily determines the position
of the CPT feature along the applied bias axis.

The dominant residual structure instead appears symmetrically on either side of the resonance and becomes more pronounced at larger magnitudes of the applied longitudinal bias. Mild differences are also visible in the detailed shape and width of the CPT feature itself. These discrepancies are particularly relevant to the estimation of \(B_y\), since the transverse field is inferred predominantly through its modification of the CPT linewidth and lineshape, rather than through a displacement of the resonance centre. Thus, even when the resonance position is reproduced well enough to constrain \(B_z\), an imperfect description of its width and surrounding spectral structure can systematically bias the inferred \(B_y\).

The bias fields selected by the adaptive search lie predominantly near the rapidly varying central region of the CPT response. This provides a plausible explanation for why the posterior can nevertheless converge to a compact estimate: the subset of the response most frequently interrogated by the adaptive trajectory is comparatively better described by the forward model response at the corresponding MAP estimates, where the experimental and simulated signals show comparatively good agreement.
The larger residuals away from this region, however, demonstrate that such local agreement does not imply global agreement of the forward model over the full \((t,B_c)\) measurement domain. The systematic calibration error so observed in \(B_y\) may therefore be due to smaller lineshape discrepancies within the informative central region, or information about the lineshape that is lost when the adaptive search samples only a restricted subset of the complete experimental response.

A natural source of such systematic disagreement is the set of physical quantities treated as fixed in the present forward model. The inference explicitly varies only $B_z$ and $B_y$, while parameters such as any small but residual $B_x$ field, probe Rabi frequency, ground-state and thermal decoherence rates, optical depth, and other experimental parameters are assigned fixed calibrated values. If any of these quantities differs appreciably from its assumed value, even within its respective order of magnitude, its effect on the transient response must be absorbed by the available inference parameters. For example, an unaccounted residual $B_x$ incorporates mixing of magnetic sublevels and has an effect of CPT contrast, while uncertainties in the decoherence and optical pumping parameters can modify the CPT linewidth and transient response of the transmitted signal. In such a case where simulation exhibits a small offset relative to the experimental data, the posterior may therefore become narrow around the best available \((B_z,B_y)\) compromise while remaining systematically displaced from the physical field values.

The likelihood scale study of Sec.~\ref{sec:sigma_dependence} provides an independent indication of the same model discrepancy. Because experiment and simulation are calibrated on the same $I_{\mathrm{sat}}$ scale, the required likelihood scale can be compared directly with the independently characterized experimental noise. Stable field estimates require values of $\sigma$ larger than the measured intensity-noise scale. Thus, the residual width required by the inference cannot be explained by experimental noise alone and must also accommodate disagreement between the experimental signal and the forward model.

The finite resolution of the precomputed \(B_y\) simulation grid and the subsequent interpolation is tested to see whether it could account for the observed bias. Replacing linear interpolation by cubic-spline interpolation produced no appreciable change in the inferred fields, making interpolation order an unlikely dominant source of the systematic discrepancy. 

The \(B_z^{\mathrm{total}}\) simulation grid is comparatively coarse (\(\Delta B_z \approx 7.7~\mu\mathrm{T}\)), so the inferred longitudinal field relies on linear interpolation between grid points. Some interpolation error is therefore expected, but the dominant systematic discrepancy in the present analysis is observed in \(B_y\), while \(B_z\) remains comparatively stable across the transverse field scans. Therefore \(B_z\) axis interpolation is not identified as a leading limitation here; if higher precision longitudinal estimation becomes necessary, finer grids or higher order interpolation can be explored.

A more complete treatment would promote the most relevant uncertain physical quantities to nuisance parameters and infer them jointly with the magnetic-field components, 
\begin{equation}
p\left(
B_z,B_y,\boldsymbol{\eta}
\mid
\mathbf{y}_{1:k}
\right),
\end{equation}
where \(\boldsymbol{\eta}\) denotes the additional forward-model parameters. Marginalizing over \(\boldsymbol{\eta}\) would propagate their uncertainty into the magnetic-field estimates rather than forcing \(B_z\) and \(B_y\) to absorb every model-data discrepancy. The computational implications of extending the present grid-based inference to such higher dimensional parameter spaces are discussed in Sec.~\ref{sec:numerical_limitations}.

As discussed in Sec.~\ref{sec:full_scan_test}, the full-scan hypothesis test favors different field values from the adaptive search estimate for several datasets, despite producing smaller global residuals. This reinforces the conclusion that different regions of the measured \((t,B_c)\) response favor different parameter compromises within the present forward model. The comparison does not establish which estimate is physically more accurate. Resolving this requires a more discriminative treatment of the target fields and relevant nuisance parameters.

\subsection{Numerical Limitations and Higher-Dimensional Inference}
\label{sec:numerical_limitations}

The present posterior is represented on a two-dimensional Cartesian grid in
$(B_z,B_y)$, which is efficient and transparent for the current problem but
scales poorly as additional parameters are introduced. Extending the inference
to include $B_x$ and marginalizing over relevant nuisance parameters would therefore make a dense tensor-product grid increasingly expensive.

Particle-based Bayesian methods, such as sequential Monte Carlo, provide a
more flexible alternative for higher dimensional inference because they avoid
explicit construction of the full Cartesian parameter grid and allow nuisance
parameters to be marginalized directly from the weighted particle ensemble
\cite{QInfer2017}. Such methods therefore provide a natural route toward
joint estimation of all three magnetic-field components together with uncertain
forward-model parameters.

\subsection{Relation to an Online Adaptive Experiment}
\label{sec:online_adaptive}

As described in Sec.~\ref{sec:offline_implementation}, the present work evaluates the Bayesian inference and adaptive measurement-selection procedure through offline replay of pre-recorded experimental data rather than through real-time control. An online implementation would additionally require sufficiently rapid posterior updating and expected information gain evaluation between successive experimental realizations, together with hardware control of the applied bias field. Such an implementation would allow the adaptive protocol to be benchmarked directly under a well-defined experimental acquisition budget, including measurements of sensitivity, dynamic range, data efficiency, and time to estimation.

\section{Conclusion}
\label{sec:conclusion}

Using pre-recorded experimental data, a joint Bayesian framework is evaluated for estimating the longitudinal and transverse magnetic-field components $(B_z,B_y)$ from the time-resolved CPT response of a cold $^{87}$Rb ensemble. The method combines a density-matrix forward model with joint Bayesian inference and an expected information gain criterion for selecting the longitudinal bias field. The present implementation is carried out entirely in post-processing and should therefore be interpreted as a proof-of-concept evaluation of the inference and measurement selection framework rather than as a demonstration of real-time adaptive magnetometry.

For the datasets considered here, the joint posterior contracts to well-localized estimates of both field components with weak posterior correlation, indicating that $B_z$ and $B_y$ are distinguishable within the adopted forward model and the measured transient response. Across independently calibrated transverse-field settings, the inferred $B_y$ tracks the applied field while the inferred $B_z$ remains comparatively stable. The transverse-field calibration nevertheless exhibits a systematic deviation from the independently calibrated values, showing that posterior concentration alone does not establish absolute field accuracy.

The likelihood scale dependence, model-data residuals, and comparison with the complete $(t,B_c)$ dataset consistently indicate that the dominant limitation of the present analysis is the fidelity of the forward model rather than the Bayesian updating procedure itself. In particular, physical quantities currently treated as fixed, including residual magnetic-field components and parameters governing optical pumping and decoherence, can modify the CPT lineshape and may therefore bias the inferred field values when their uncertainty is not included in the inference.

A natural extension is to infer all three magnetic-field components and also marginalize over the relevant nuisance parameters. Because the computational cost of the present grid representation grows rapidly with parameter dimension, particle-based methods such as sequential Monte Carlo provide a practical route toward such higher dimensional inference.

% The present study does not establish an adaptive improvement in sensitivity, dynamic range, measurement time, or data efficiency. Demonstrating such figures of merit requires an online implementation in which posterior updating, information-gain evaluation, and experimental control are performed between successive measurement realizations under a well-defined acquisition budget.
% The offline results presented here establish the feasibility of using posterior dependent measurement selection for joint field inference using the transient CPT response and identify the principal modeling and computational requirements for a future real-time implementation.
The results presented here establish the feasibility of using posterior dependent measurement selection to optimize the information gained from the the transient CPT response toward joint field inference and identify the principal modeling and computational requirements for a future real-time implementation.

\section*{Acknowledgments}
This work was supported by the National Quantum Mission of the Department of Science and Technology (DST), Government of India. S.J. acknowledges financial support from the Ministry of Human Resource Development (MHRD), Government of India. G.M. acknowledges support from the Prime Minister’s Research Fellowship (PMRF). The authors also acknowledge the use of Claude (Anthropic) for assistance in improving the clarity and quality of the manuscript.

\clearpage
\appendix

\section{Full-Scan Hypothesis-Test Results}
\label{app:full_scan_test_appendix}
% \begin{figure}[t]
%     \centering
%     \includegraphics[width=\columnwidth]{Figures/Joint_Estimation/Crude_Estimate_Calibration.pdf}
%     \caption{Transverse field calibration of the maximizing full-scan hypotheses. The solid line is a linear fit to $\hat{B}^*_y$ versus $B_{y,\mathrm{FID}}$, with slope $1.433\pm0.043$ and $R^2=0.998$; the dashed line denotes the ideal relation $\hat{B}^*_y=B_{y,\mathrm{FID}}$. Only $\hat{B}^*_y$ is shown because the maximizing $\hat{B}^*_z$ is the same discrete grid value for all datasets.}
%     \label{fig:full_scan_calibration}
% \end{figure}

% \begin{figure}[ht]
%     \centering

%     \includegraphics[width=\linewidth]{Figures/Joint_Estimation/By_2_Difference_Plot_Between_Experiment_and_Simulation_based_on_Full_Scan_Estimate.png}

%     \caption{Absolute residual between the experimentally measured probe transmission and the forward-model prediction evaluated at the maximizing full-scan hypothesis for the $B_{y,\mathrm{FID}}=35~\mu\mathrm{T}$ dataset. The summed squared residual over the displayed $(t,B_c)$ domain is $L=0.88 \, I^2_{\mathrm{sat}}$}
%     \label{fig:difference_plots_full_scan}
% \end{figure}

\begin{figure}[t]
    \centering
    \includegraphics[width=\columnwidth]{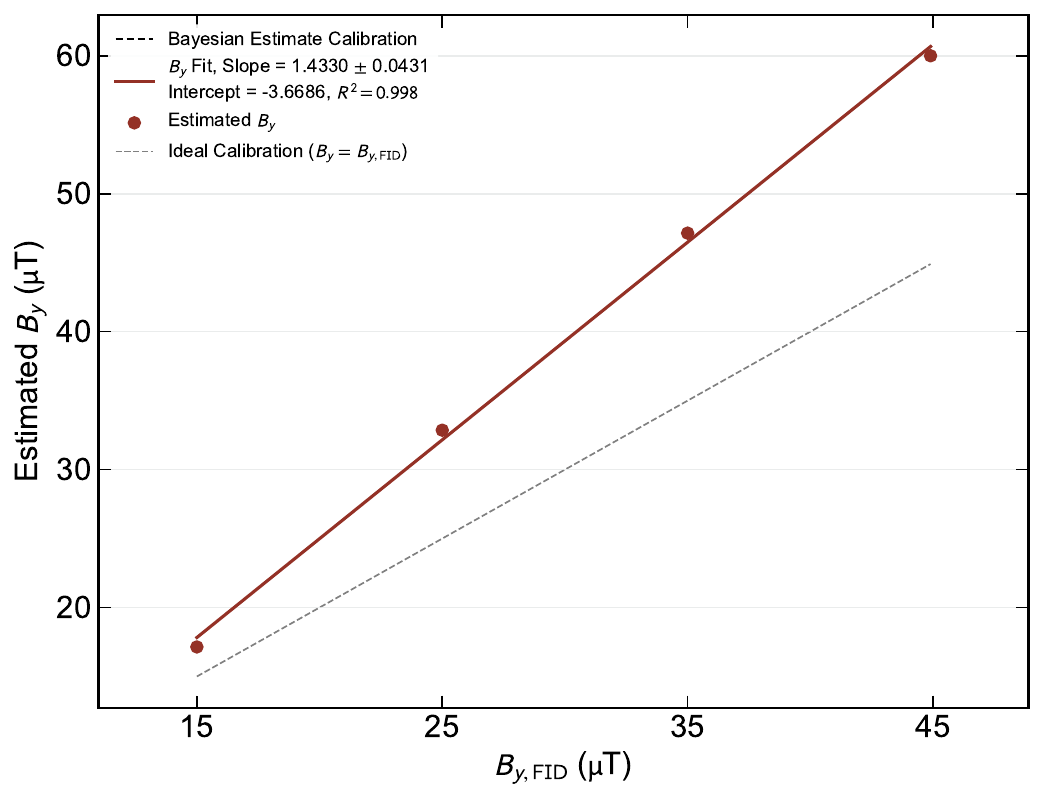}
    \caption{Transverse field calibration of the maximizing full-scan hypotheses. The solid line is a linear fit to $\hat{B}^*_y$ versus $B_{y,\mathrm{FID}}$, with slope $1.433\pm0.043$ and $R^2=0.998$; the dashed line denotes the ideal relation $\hat{B}^*_y=B_{y,\mathrm{FID}}$. Only $\hat{B}^*_y$ is shown because the maximizing $\hat{B}^*_z$ is the same discrete grid value for all datasets.}
    \label{fig:full_scan_calibration}
\end{figure}

As shown in Fig.~\ref{fig:full_scan_calibration}, the full-scan estimates vary nearly linearly with the independently calibrated $B_{y,\mathrm{FID}}$, but the fitted
slope,
\begin{equation}
    B_y^\star
    =
    (1.433 \pm 0.043)\,B_{y,\mathrm{FID}}
    - 3.669~\mu\mathrm{T},
\end{equation}
differs substantially from the ideal unit slope, despite the good linear fit ($R^2=0.998$). Thus, the full-scan test also exhibits a systematic
transverse field calibration error.

\begin{figure}[ht]
    \centering

    \includegraphics[width=\linewidth]{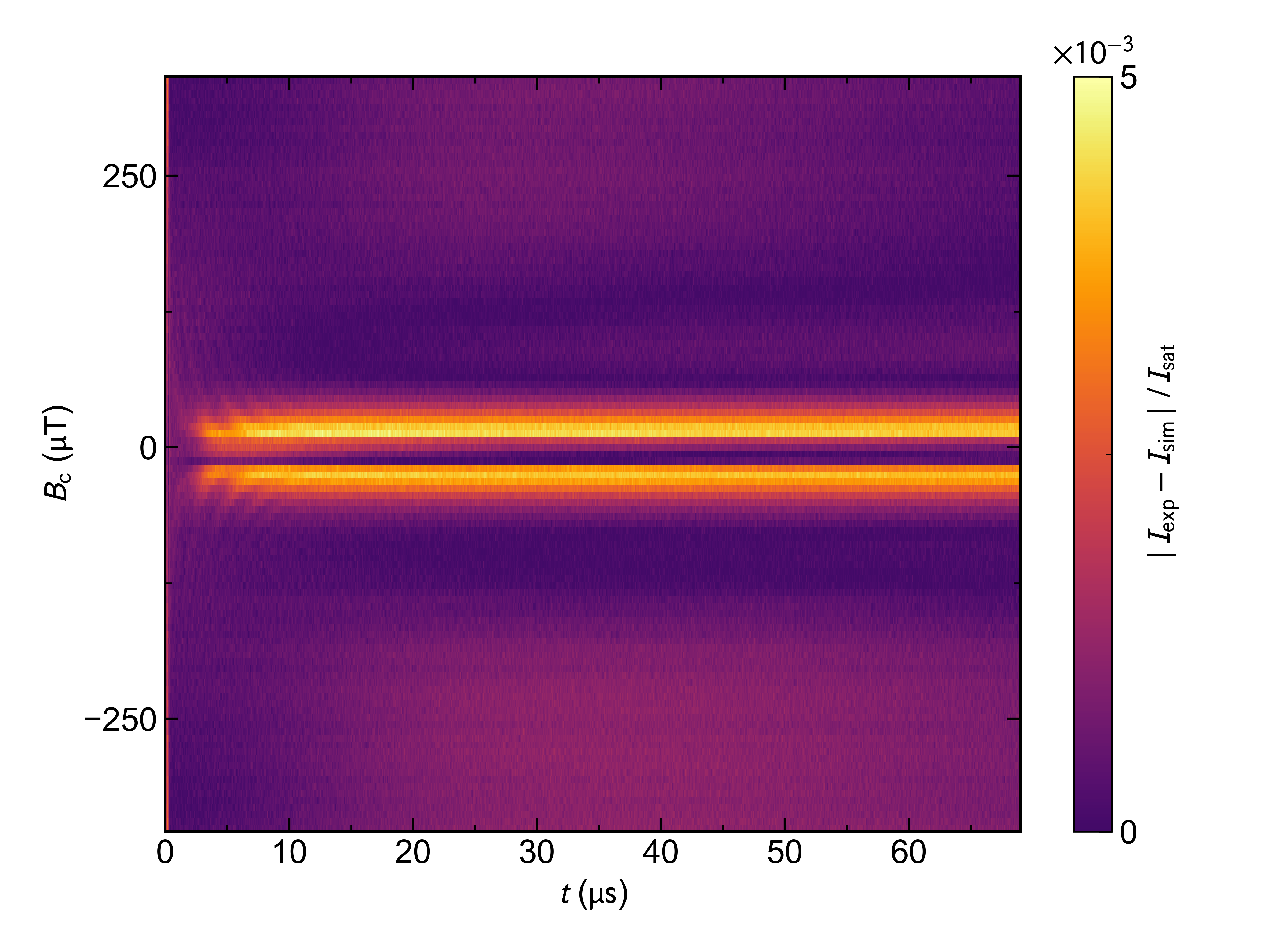}

    \caption{Absolute residual between the experimentally measured probe transmission and the forward-model prediction evaluated at the maximizing full-scan hypothesis for the $B_{y,\mathrm{FID}}=35~\mu\mathrm{T}$ dataset. The summed squared residual over the displayed $(t,B_c)$ domain is $L=0.88 \, I^2_{\mathrm{sat}}$}
    \label{fig:difference_plots_full_scan}
\end{figure}

The residual maps evaluated at the maximizing full-scan hypotheses are shown
in Fig.~\ref{fig:difference_plots_full_scan}. Their aggregate residuals are lower than
those obtained from the adaptive Bayesian estimates, as expected because the
full-scan hypotheses are selected using the complete measured response. The
broad residuals at large $|B_c|$ are reduced, while structured discrepancies
remain around the CPT feature, particularly in its width and surrounding
lineshape.

The lower global residual therefore reflects a different best-fit compromise
within the present forward model rather than removal of the model discrepancy.
Since both the adaptive and full-scan procedures retain non-ideal $B_y$
calibration, this comparison does not determine which estimate is closer to
the physical transverse field.

\clearpage

\bibliography{references}

\end{document}